\documentclass[english,prb,aps,preprintnumbers,amsmath,amssymb,twocolumn]{revtex4}
\usepackage[T1]{fontenc}
\usepackage[latin9]{inputenc}
\usepackage{amsmath}
\usepackage{amssymb}
\usepackage{graphicx}
\usepackage{esint}

\makeatletter
\@ifundefined{textcolor}{}
{%
 \definecolor{BLACK}{gray}{0}
 \definecolor{WHITE}{gray}{1}
 \definecolor{RED}{rgb}{1,0,0}
 \definecolor{GREEN}{rgb}{0,1,0}
 \definecolor{BLUE}{rgb}{0,0,1}
 \definecolor{CYAN}{cmyk}{1,0,0,0}
 \definecolor{MAGENTA}{cmyk}{0,1,0,0}
 \definecolor{YELLOW}{cmyk}{0,0,1,0}
}

\usepackage{float}\usepackage{subfigure}\usepackage{dcolumn}
\usepackage{bm}\usepackage{enumerate}\usepackage{babel}\usepackage{hyperref}

\makeatother

\usepackage{babel}
\begin{document}

\title{Symmetry breaking and restoration using the equation-of-motion technique
for nonequilibrium quantum impurity models}

\author{Tal J. Levy and Eran Rabani}

\affiliation{School of Chemistry, The Sackler Faculty of Exact Sciences, Tel Aviv
University, Tel Aviv 69978, Israel}
\begin{abstract}
The description of the dynamics of correlated electrons in quantum
impurity models is typically described within the nonequilibrium Green
function formalism combined with a suitable approximation. One common
approach is based on the equation-of-motion technique often used to
describe different regimes of the dynamic response. Here, we show
that this approach may violate certain symmetry relations that must
be fulfilled by the definition of the Green functions. These broken
symmetries can lead to unphysical behavior. To circumvent this pathological
shortcoming of the equation-of-motion approach we provide a scheme
to restore basic symmetry relations. Illustrations are given for the
Anderson and double Anderson impurity models. 
\end{abstract}
\maketitle

\section{Introduction\label{sec:Introduction}}

Describing the transport of electrons through an interacting region
is a challenging task and typically involves the calculation of the
dynamics of correlated electrons driven away from equilibrium~\citealp{Datta1995,Datta2005,Imry_book}.
In general, this many body out-of-equilibrium problem cannot be solved
exactly but for a few simple cases~\citealp{Jauho1994,Schiller1995,Wang1996,Swenson2011}.
Excluding recent developments based on brute-force approaches such
as time-dependent numerical renormalization-group techniques~\citealp{anders_real-time_2005,schmitteckert_nonequilibrium_2004,white_density_1992},
iterative~\citealp{weiss_iterative_2008,eckel_comparative_2010,Segal10}
or stochastic~\citealp{muehlbacher_real-time_2008,werner_diagrammatic_2009,werner_weak-coupling_2010,schiro_real-time_2009,gull10:_bold_monte_carlo}
diagrammatic techniques to real time path integral formulations, wave
function based approaches~\citealp{Thoss2011}, or reduced dynamic
approaches~\citealp{leijnse_kinetic_2008,Cohen2011}, all suitable
to relatively simple model systems, most theoretical treatments of
quantum transport rely on approximations of some sort. One well studied
approach is based on the nonequilibrium Green function (NEGF) formalism
otherwise known as the Keldysh NEGF or the Schwinger-Keldysh formalism~\citealp{Schwinger1961a,Keldysh1964},
which is widely used to describe transport phenomena~\citealp{Ratner2007,Xue2002,Datta2000}. 

Based on the NEGF, an exact expression for the stationary current
through an interacting system coupled to large non-interacting metallic
leads in terms of the system's Green function can be derived~\citealp{Meir1992}:
\begin{eqnarray}
I & = & \frac{ie}{2\pi\hbar}\int\mbox{d}\varepsilon\left(\mbox{Tr}\left\{ f_{L}\left(\varepsilon-\mu_{L}\right)\boldsymbol{\Gamma}_{L}\left(\varepsilon\right)\right.\right.\nonumber \\
 &  & \times\left.\left.\left(\mathbf{G}^{r}\left(\varepsilon\right)-\mathbf{G}^{a}\left(\varepsilon\right)\right)\right\} +\mbox{Tr}\left\{ \boldsymbol{\Gamma}_{L}\mathbf{G}^{<}\left(\varepsilon\right)\right\} \right),\label{eq:I1}
\end{eqnarray}
or equivalently 
\begin{eqnarray}
I & = & \frac{e}{h}\int\mbox{d}\varepsilon\mbox{Tr}\left\{ \boldsymbol{\Sigma}_{L}^{<}\left(\varepsilon\right)\mathbf{G}^{>}\left(\varepsilon\right)-\boldsymbol{\Sigma}_{L}^{>}\left(\varepsilon\right)\mathbf{G}^{<}\left(\varepsilon\right)\right\} \label{eq:I2}
\end{eqnarray}
where $\mathbf{G}^{r}$ ($\mathbf{G}^{a}$) is the retarded (advanced)
Green function (GF) of the system, $\mathbf{G}^{<}$ ($\mathbf{G}^{>}$)
is the lesser (greater) GF of the system, which will be defined later
below. The lesser tunneling self-energy is given by $\boldsymbol{\Sigma}_{L0}^{<}=if_{L}\left(\varepsilon-\mu_{L}\right)\boldsymbol{\Gamma}_{L}$,
where $f_{k}\left(\varepsilon-\mu_{k}\right)$ is the Fermi\textendash{}Dirac
distribution and $\boldsymbol{\Gamma}_{L}$ is the matrix coupling
the interacting system to the left reservoir with elements $\left(\Gamma_{L}\right)_{mn}=2\pi\rho_{k}\left(\varepsilon\right)t_{kn}t_{km}^{*}$
($t_{km}$ is the hopping matrix elements between the system and the
left reservoir). The calculation of the system's GF required to obtain
the current (or other observables) is far from trivial, excluding
simple noninteracting cases. Most applications are based on perturbative
diagrammatic techniques to obtain $\mathbf{G}^{r}$, $\mathbf{G}^{a}$,
$\mathbf{G}^{<}$ and $\mathbf{G}^{>}$~\citealp{Haug1996}. Alternatively,
one can use the equation-of-motion (EOM) approach, which allows to
deduce the system's GFs by deriving the corresponding equations of
motion~\citealp{Zubarev1960,Lacroix1981,Scheck2007}. In light of
its simplicity, it has been used extensively to describe transport
phenomena such as the Coulomb blockade~\citealp{Song2007} and the
Kondo effect~\citealp{Lacroix1981,Meir1991,Galperin2007a}, providing
qualitative and in some cases quantitative results. When applied to
interacting systems, the EOM for the GF gives rise to an infinite
hierarchy of equations of higher-order GFs. A well-known approximation
procedure is then to truncate this hierarchy, thus introducing a mean-field
like description to some observables. These equations for the GFs
then need to be solved self-consistently for the resulting closed
set of equations. Although successful, the EOM technique has its drawbacks~\citealp{Kashcheyevs2006}. 

In this paper we show that while a closure can always be obtained,
it is not clear a priori whether it fulfills symmetry relations that
single particle GFs must obey. This failure can lead to solutions
which are not physical, such as complex occupation of levels and even
finite currents at zero bias. We also propose an approach to fix this
deficiency by imposing a set of rules to reconstruct GFs that fulfill
basic symmetry relations. Illustrations are given for the Anderson
model~\citealp{Anderson1961} at the Kondo regime and for the double
Anderson model~\citealp{Jayaprakash81}. Our paper is organized as
follows: in Sec. \ref{sec:nonequilibrium-EOM-technique} we describe
the EOM approach and the single site and double site Anderson models.
In Sec. \ref{sec:Symmetry-Breaking} we discuss symmetry relation
for GFs and illustrate symmetry breaking for the aforementioned models
with specific closures suitable to describe the Kondo effect. In Sec.
\ref{sec:Symmetry-restoration} we provide a recipe to restore the
basic symmetry relations within the EOM approach and discuss implications
for level occupancy and coherences, current, and sum rules for the
Anderson model in the Kondo regime and the double Anderson model.
Finally, in Sec. \ref{sec:Conclusions-and-Summary} we conclude.

\section{EOM technique and models\label{sec:nonequilibrium-EOM-technique}}

\subsection{Equations of motion\label{sub:Equations-of-motion}}

The EOM for the contour ordered GF~\citealp{Marques2006} is obtained
from the Heisenberg EOM for a Heisenberg operator $\frac{\mbox{d}}{\mbox{d}t}\hat{A}\left(t\right)=\frac{i}{\hbar}\left[\hat{H}\left(t\right),\hat{A}_{H}\left(t\right)\right]+\frac{\partial}{\partial t}\hat{A}_{H}\left(t\right),$
where in our case $\hat{H}\left(t\right)=\hat{H}_{0}+\hat{V}\left(t\right)$.
Here $\hat{H}_{0}$ stands for the one body noninteracting part of
$\hat{H}\left(t\right)$, $\hat{V}\left(t\right)=\hat{H}\left(t\right)-\hat{H}_{0}$,
and $\left[\hat{A},\hat{B}\right]$ is the commutator. Let us consider
a generic example. We define the contour ordered GF 
\begin{equation}
G\left(\mathbf{r}_{2},t_{2},\mathbf{r}_{1},t_{1}\right)=-\frac{i}{\hbar}\left\langle T_{C}\hat{\Psi}_{H}\left(\mathbf{r}_{2},t_{2}\right)\hat{\Psi}_{H}^{\dagger}\left(\mathbf{r}_{1},t_{1}\right)\right\rangle ,
\end{equation}
where $T_{C}$ is the contour time ordering operator and $\hat{\Psi}_{H}\,\left(\hat{\Psi}_{H}^{\dagger}\right)$
is the system's annihilation (creation) field operator in the Heisenberg
picture (in what follows we omit the $H$ index). The EOM~\citealp{Niu1999a}
for $G\left(\mathbf{r}_{2},t_{2},\mathbf{r}_{1},t_{1}\right)$ can
be written as (omitting the $\mathbf{r}$ dependence for brevity)
\begin{eqnarray}
G\left(t_{2},t_{1}\right) & = & g_{2}\left(t_{2},t_{1}\right)\left\langle \left\{ \hat{\Psi},\hat{\Psi}^{\dagger}\right\} \right\rangle \nonumber \\
 &  & -\frac{i}{\hbar}\int_{C}\mbox{d}t\, g_{2}\left(t_{2},t\right)\label{eq:contourGF}\\
 &  & \times\left\langle T_{C}\left[\hat{\Psi}\left(t\right),\hat{V}\left(t\right)\right]\hat{\Psi}^{\dagger}\left(t_{1}\right)\right\rangle ,\nonumber 
\end{eqnarray}
where $\left(i\hbar\frac{\partial}{\partial t_{2}}-\varepsilon\right)g_{2}\left(t_{2},t_{1}\right)=\delta\left(t_{1}-t_{2}\right),$
$\left\{ \hat{A},\hat{B}\right\} $ is the anti-commutator, and $\varepsilon$
is defined from the equation, $\varepsilon\hat{\Psi}\left(t\right)=\left[\hat{\Psi}\left(t\right),\hat{H}_{0}\right].$
For example, if $\hat{H}_{0}=\sum_{n}\left(\varepsilon_{n}-\mu\right)\hat{d}_{n}^{\dagger}\hat{d}_{n}$
and $\hat{\Psi}=\hat{d}_{i}$ then $\varepsilon=\varepsilon_{i}-\mu$.
Following Langreth theorem~\citealp{Langreth76}, we can change the
contour integration in equation (\ref{eq:contourGF}) to integration
along the real time axis. This yields (see Sec. \ref{sec:Symmetry-Breaking}
for the definitions of the different real-time GFs) 
\begin{eqnarray}
G^{r}\left(t_{2},t_{1}\right) & = & g_{2}^{r}\left(t_{2},t_{1}\right)\left\langle \left\{ \hat{\Psi},\hat{\Psi}^{\dagger}\right\} \right\rangle \nonumber \\
 &  & +\int_{t_{1}}^{t_{2}}\mbox{d}t\, g_{2}^{r}\left(t_{2},t\right)\mathbb{G}^{r}\left(t,t_{1}\right),\nonumber \\
G^{<}\left(t_{2},t_{1}\right) & = & g_{2}^{<}\left(t_{2},t_{1}\right)\left\langle \left\{ \hat{\Psi},\hat{\Psi}^{\dagger}\right\} \right\rangle \\
 &  & +\int_{t_{0}}^{t_{2}}\mbox{d}t\, g_{2}^{r}\left(t_{2},t\right)\mathbb{G}^{<}\left(t,t_{1}\right)\nonumber \\
 &  & +\int_{t_{0}}^{t_{1}}\mbox{d}t\, g_{2}^{<}\left(t_{2},t\right)\mathbb{G}^{a}\left(t,t_{1}\right),\nonumber 
\end{eqnarray}
where $G^{r}\left(t_{2},t_{1}\right)$ is the retarded GF usually
used to calculate the response of the system at time $t_{2}$ to an
earlier perturbation of the system at time $t_{1}$. $G^{<}\left(t_{2},t_{1}\right)$
is the lesser GF which plays the role of the single particle density
matrix, and $\mathbb{G}\left(t_{2},t_{1}\right)=-\frac{i}{\hbar}\left\langle T_{C}\left[\hat{\Psi}\left(t_{2}\right),\hat{V}\left(t_{2}\right)\right]\hat{\Psi}^{\dagger}\left(t_{1}\right)\right\rangle $
is a new GF generated by the EOM procedure. Depending on the Hamiltonian
it can be a single particle GF or a many particle GF and can involve
lead operators as well as system operators. In steady state, the GFs
depend only on the difference in time, $t=t_{2}-t_{1}$, which is
simpler to express in Fourier space 
\begin{eqnarray}
G^{r}\left(\omega\right) & = & g_{2}^{r}\left(\omega\right)\left\langle \left\{ \hat{\Psi},\hat{\Psi}^{\dagger}\right\} \right\rangle +g_{2}^{r}\left(\omega\right)\mathbb{G}^{r}\left(\omega\right),\\
G^{<}\left(\omega\right) & = & g_{2}^{<}\left(\omega\right)\left\langle \left\{ \hat{\Psi},\hat{\Psi}^{\dagger}\right\} \right\rangle +g_{2}^{r}\left(\omega\right)\mathbb{G}^{<}\left(\omega\right)\nonumber \\
 &  & +g_{2}^{<}\left(\omega\right)\mathbb{G}^{a}\left(\omega\right).
\end{eqnarray}
To simplify the notation we denote the Fourier transform of $G\left(t_{2}-t_{1}\right)=G\left(t\right)$
as $G\left(\omega\right)$, i.e., functions with an argument ``$\omega$''
are Fourier transforms of their time-domain counterparts. At this
stage one has to evaluate $\mathbb{G}\left(t_{2},t_{1}\right)$ ($\mathbb{G}\left(t\right)$
in steady state). Except for very simple cases, where an exact closure
can be obtained, writing the EOM for $\mathbb{G}\left(t_{2},t_{1}\right)$
will produce new and/or ``higher order'' GFs that need to be evaluated.
This leads (in principle) to an infinite set of equations. The idea
of the EOM method is therefore, to truncate this hierarchy of equations
making a mean-field like approximation for the ``higher-order''
GFs through lower order functions. This is the Achilles heel of this
method as there is no systematic way to close the equations. Usually
the approximations have physical meaning within the regime of the
problem at hand~\citealp{Plakida1970a,Ihle1973,Meir1991}. In what
follows we demonstrate that different approximations can sometimes
break symmetry relations that the GFs must fulfill. We will use two
impurity models to demonstrate at what level of approximation the
symmetry relations are violated and propose a scheme to restore symmetrization.

\subsection{The impurity models\label{sub:The-models}}

To illustrate the shortcomings of the EOM approach, we refer to the
Anderson model~\citealp{Anderson1961,Meir1991,Wingreen1994} and
the double Anderson model~\citealp{Jayaprakash81} to represent two
different degrees of complexity in correlated systems. As commonly
used, we split the total Hamiltonian into three parts~\citealp{Haug1996}:
\begin{equation}
\hat{H}=\hat{H}_{sys}+\hat{H}_{bath}+\hat{H}_{int},
\end{equation}
where $\hat{H}_{bath}$ describes the macroscopic leads (left and
right contacts), $\hat{H}_{sys}$ describes the system of interest
(in our case the impurities), and $\hat{H}_{int}$ is the interaction
Hamiltonian between the system and the leads. The contacts (leads)
are modeled as infinite non-interacting fermionic baths~\citealp{Meirav1990,Altshuler1991,Wees1998}
with a Hamiltonian in second quantization given by 
\begin{equation}
\hat{H}_{bath}=\sum_{\sigma,k\in\{L,R\}}\epsilon_{k,\sigma}c_{k,\sigma}^{\dagger}c_{k,\sigma},
\end{equation}
where $\epsilon_{k,\sigma}$ is the energy of a free electron in the
left ($L$) or right ($R$) lead, in momentum state $k$ and spin
$\sigma$. The operator $c_{k,\sigma}\,\left(c_{k,\sigma}^{\dagger}\right)$
is the annihilation (creation) operator of such an electron. The form
chosen for $\hat{H}_{sys}$ depends on the system studied. For the
Anderson impurity model~\citealp{Anderson1961} 
\begin{equation}
\hat{H}_{sys}=\sum_{\sigma\in\{\uparrow,\downarrow\}}\epsilon_{\sigma}n_{\sigma}+Un_{\uparrow}n_{\downarrow}.
\end{equation}
Here $n_{\sigma}=d_{\sigma}^{\dagger}d_{\sigma}$ is the number operator
of the spin $\sigma$ electron with energy $\varepsilon_{\sigma}$
and $U$ is the repulsion energy between two electrons on the same
site with opposite spins (intra-site repulsion). The second model
we discuss is the double Anderson model~\citealp{Jayaprakash81}
\begin{eqnarray}
\hat{H}_{sys} & = & \sum_{\sigma,m\in\{\alpha,\beta\}}\epsilon_{m\sigma}n_{m\sigma}+\sum_{m}U_{m}n_{m\uparrow}n_{m\downarrow}\\
 &  & +\sum_{\sigma,\sigma'}V_{\alpha\beta}^{\sigma\sigma'}n_{\alpha\sigma}n_{\beta\sigma'}+\sum_{\sigma}\left[h_{\alpha\beta}^{\sigma}d_{\alpha}^{\dagger}d_{\beta}+h.c.\right],\nonumber 
\end{eqnarray}
where the first two terms on the R.H.S are similar to the Anderson
impurity model Hamiltonian (extended to $2$ sites), $V_{\alpha\beta}^{\sigma\sigma'}$
is the repulsion energy between two electrons on different sites (inter-site
repulsion), and $h_{\alpha\beta}^{\sigma}$ is the coupling strength
for electron hopping between the two sites. The interaction between
the system and the contacts is simply given by the tunneling Hamiltonian~\citealp{Caroli1971}
\begin{equation}
\hat{H}_{int}=\sum_{m,\sigma,k\in\{L,R\}}t_{k,m}^{\sigma}c_{k,\sigma}^{\dagger}d_{m,\sigma}+h.c..
\end{equation}
The parameter $t_{k,m}^{\sigma}$ represents the coupling strength
between the system and the leads, and the index $m$ runs over the
site index $\left\{ \alpha,\beta\right\} $ in the double Anderson
model.

\section{symmetry breaking in impurity models\label{sec:Symmetry-Breaking}}

\subsection{Definitions and symmetry relations\label{sub:Definitions-and-symmetry}}

In the Keldysh formalism the two time NEGF is defined on a contour.
In accordance with where on the contour the two times are placed one
can define six real-time GFs~\citealp{Mahan1990}; the time-ordered
$G^{t}$, anti-time ordered $G^{\bar{t}}$, lesser $G^{<}$, greater
$G^{>}$, retarded $G^{r}$, and advanced $G^{a}$:
\begin{eqnarray}
G_{\alpha\beta}^{t}\left(t_{2},t_{1}\right) & = & -\frac{i}{\hbar}\theta\left(t_{2}-t_{1}\right)\left\langle \hat{\Psi}_{\alpha}\left(t_{2}\right)\hat{\Psi}_{\beta}^{\dagger}\left(t_{1}\right)\right\rangle \nonumber \\
 &  & +\frac{i}{\hbar}\theta\left(t_{1}-t_{2}\right)\left\langle \hat{\Psi}_{\beta}^{\dagger}\left(t_{1}\right)\hat{\Psi}_{\alpha}\left(t_{2}\right)\right\rangle ,\nonumber \\
G_{\alpha\beta}^{\bar{t}}\left(t_{2},t_{1}\right) & = & -\frac{i}{\hbar}\theta\left(t_{1}-t_{2}\right)\left\langle \hat{\Psi}_{\alpha}\left(t_{2}\right)\hat{\Psi}_{\beta}^{\dagger}\left(t_{1}\right)\right\rangle \nonumber \\
 &  & +\frac{i}{\hbar}\theta\left(t_{2}-t_{1}\right)\left\langle \hat{\Psi}_{\beta}^{\dagger}\left(t_{1}\right)\hat{\Psi}_{\alpha}\left(t_{2}\right)\right\rangle ,\nonumber \\
G_{\alpha\beta}^{<}\left(t_{2},t_{1}\right) & = & \frac{i}{\hbar}\left\langle \hat{\Psi}_{\beta}^{\dagger}\left(t_{1}\right)\hat{\Psi}_{\alpha}\left(t_{2}\right)\right\rangle ,\\
G_{\alpha\beta}^{>}\left(t_{2},t_{1}\right) & = & -\frac{i}{\hbar}\left\langle \hat{\Psi}_{\alpha}\left(t_{2}\right)\hat{\Psi}_{\beta}^{\dagger}\left(t_{1}\right)\right\rangle ,\nonumber \\
G_{\alpha\beta}^{r}\left(t_{2},t_{1}\right) & = & -\frac{i}{\hbar}\theta\left(t_{2}-t_{1}\right)\left\langle \left\{ \hat{\Psi}_{\alpha}\left(t_{2}\right),\hat{\Psi}_{\beta}^{\dagger}\left(t_{1}\right)\right\} \right\rangle ,\nonumber \\
G_{\alpha\beta}^{a}\left(t_{2},t_{1}\right) & = & \frac{i}{\hbar}\theta\left(t_{1}-t_{2}\right)\left\langle \left\{ \hat{\Psi}_{\alpha}\left(t_{2}\right),\hat{\Psi}_{\beta}^{\dagger}\left(t_{1}\right)\right\} \right\rangle .\nonumber 
\end{eqnarray}
The retarded GF can be used to calculate the response of the system
at time $t_{2}$ to an earlier perturbation of the system at time
$t_{1}$ and is proportional to the local density of states, while
the lesser GF is also known as the particle propagator and plays the
role of the single particle density matrix. From equation (\ref{eq:I1})
it is obvious that in order to calculate the stationary current the
retarded, advanced and lesser GFs are needed, thus, the current is
expressed in terms of the local density of states and the occupation
of the system. Using the given definitions it is clear that the following
relations must hold:
\begin{eqnarray}
G_{\alpha\beta}^{r}\left(t_{2},t_{1}\right) & = & \left(G_{\beta\alpha}^{a}\left(t_{1},t_{2}\right)\right)^{*},\nonumber \\
G_{\alpha\beta}^{<,>}\left(t_{2},t_{1}\right) & = & -\left(G_{\beta\alpha}^{<,>}\left(t_{1},t_{2}\right)\right)^{*},\label{eq:timeRelations}\\
G_{\alpha\beta}^{r}\left(t_{2},t_{1}\right)-G_{\alpha\beta}^{a}\left(t_{2},t_{1}\right) & = & G_{\alpha\beta}^{>}\left(t_{2},t_{1}\right)-G_{\alpha\beta}^{<}\left(t_{2},t_{1}\right).\nonumber 
\end{eqnarray}
In steady state these relations can be rewritten in Fourier space
as:
\begin{eqnarray}
G_{\alpha\beta}^{r}\left(\omega\right) & = & \left(G_{\beta\alpha}^{a}\left(\omega\right)\right)^{*},\nonumber \\
G_{\alpha\beta}^{<,>}\left(\omega\right) & = & -\left(G_{\beta\alpha}^{<,>}\left(\omega\right)\right)^{*},\label{eq:relations}\\
G_{\alpha\beta}^{r}\left(\omega\right)-G_{\alpha\beta}^{a}\left(\omega\right) & = & G_{\alpha\beta}^{>}\left(\omega\right)-G_{\alpha\beta}^{<}\left(\omega\right).\nonumber 
\end{eqnarray}
In what follows we show that these relations do not hold when the
GFs are obtained by the EOM technique with an arbitrary closure.

\subsection{The Anderson model\label{sub:The-Anderson-model}}

Following the derivation in Refs.~\onlinecite{Caroli1971,Meir1991,Haug1996}
we define the following contour ordered GF:
\begin{equation}
G_{\sigma\sigma}\left(t,t'\right)=-\frac{i}{\hbar}\left\langle T_{C}d_{\sigma}\left(t\right)d_{\sigma}^{\dagger}\left(t'\right)\right\rangle ,
\end{equation}
\begin{equation}
G_{2}\left(t,t'\right)=-\frac{i}{\hbar}\left\langle T_{C}n_{\bar{\sigma}}\left(t\right)d_{\sigma}\left(t\right)d_{\sigma}^{\dagger}\left(t'\right)\right\rangle ,
\end{equation}
where $\bar{\sigma}$ is the opposite spin of $\sigma$. Various approximate
decoupling procedures can be applied to the many particle GF~\citealp{Pals1996}.
Here we follow the approximation scheme used in Refs.~\onlinecite{Meir1991,Haug1996}
where all electronic correlations containing at most one lead operator,
are not decoupled and their EOM are calculated. Higher order GFs involving
(opposite) spin correlations in the leads are set to zero, and the
remaining higher order GFs involving lead and system degrees of freedom
are decoupled such that $F_{2}\left(t,t'\right)=-\frac{i}{\hbar}\left\langle T_{C}c_{kn\bar{\sigma}}^{\dagger}\left(t\right)d_{\sigma}\left(t\right)c_{qm\bar{\sigma}}\left(t\right)d_{\sigma}^{\dagger}\left(t'\right)\right\rangle =-\delta_{kq}\delta_{mn}f_{k}\left(\varepsilon_{n}-\mu_{k}\right)G_{\sigma\sigma}\left(t,t'\right)$.
The resulting EOMs (in Fourier space) are: 
\begin{equation}
\left(\hbar\omega-\varepsilon_{\sigma}-\Sigma_{0}\left(\omega\right)\right)G_{\sigma\sigma}\left(\omega\right)=1+UG_{2}\left(\omega\right),\label{eq:And1}
\end{equation}
\begin{eqnarray}
G_{2}\left(\omega\right) & = & \left(\hbar\omega-\varepsilon_{\sigma}-U-\Sigma_{0}\left(\omega\right)-\Sigma_{3}\left(\omega\right)\right)^{-1}\nonumber \\
 &  & \times\left(\left\langle n_{\bar{\sigma}}\right\rangle -\Sigma_{1}\left(\omega\right)G_{\sigma\sigma}\left(\omega\right)\right),\label{eq:And2}
\end{eqnarray}
where $\left\langle n_{\bar{\sigma}}\right\rangle =-\frac{i\hbar}{2\pi}\int_{-\infty}^{\infty}G_{\bar{\sigma}\bar{\sigma}}^{<}\left(\omega\right)\mbox{d}\omega$,
$\Sigma_{0}\left(\omega\right)=\sum_{i,k\in\left\{ L,R\right\} }\frac{\left|t_{k\sigma}\right|^{2}}{\hbar\omega-\varepsilon_{k,i,\sigma}}$
is the exact self-energy for the non-interacting case, $\Sigma_{1}\left(\omega\right)$
and $\Sigma_{3}\left(\omega\right)$ are the self-energies due to
the tunneling of the $\bar{\sigma}$ electron , and are given by
\begin{eqnarray}
\Sigma_{j}\left(\omega\right) & = & \sum_{k\in\left\{ L,R\right\} }A_{k}^{\left(j\right)}\left|t_{k\sigma}\right|^{2}\nonumber \\
 &  & \times\left(\frac{1}{\hbar\omega+\varepsilon_{k,\bar{\sigma}}-\varepsilon_{\sigma}-\varepsilon_{\bar{\sigma}}-U}\right.\\
 &  & +\left.\frac{1}{\hbar\omega-\varepsilon_{k,\bar{\sigma}}-\varepsilon_{\sigma}+\varepsilon_{\bar{\sigma}}}\right),\, j=1,3\nonumber 
\end{eqnarray}
with $A_{k}^{\left(1\right)}=f_{k}\left(\varepsilon_{k,\sigma}-\mu_{k}\right)$,
$A_{k}^{\left(3\right)}=1$, and $f_{k}\left(\varepsilon_{k,\sigma}-\mu_{k}\right)$
is the Fermi Dirac distribution. To show that these set of equations
break the symmetry relation $G_{\sigma\sigma}^{<}\left(\omega\right)=-\left(G_{\sigma\sigma}^{<}\left(\omega\right)\right)^{*}$
we define 
\begin{equation}
\Sigma_{4}\left(\omega\right)=\Sigma_{0}\left(\omega\right)+\Sigma_{3}\left(\omega\right),
\end{equation}
\begin{equation}
g\left(\omega\right)=\frac{1}{\hbar\omega-\varepsilon_{\sigma}-\Sigma_{0}\left(\omega\right)},
\end{equation}
\begin{equation}
g_{2}\left(\omega\right)=\frac{1}{\hbar\omega-\varepsilon_{\sigma}-U-\Sigma_{4}\left(\omega\right)}.
\end{equation}
With these definitions equations (\ref{eq:And1}) and (\ref{eq:And2})
can be rewritten (omitting $\left(\omega\right)$ for brevity) as:
\begin{equation}
G_{\sigma\sigma}=g+gUG_{2},\label{eq:AndG}
\end{equation}
\begin{equation}
G_{2}=g_{2}\left\langle n_{\bar{\sigma}}\right\rangle -g_{2}\Sigma_{1}G_{\sigma\sigma}.\label{eq:AndG2}
\end{equation}
Substituting equation (\ref{eq:AndG2}) in equation (\ref{eq:AndG})
and applying the Langreth rules we find that the lesser GF is given
by 
\begin{eqnarray}
G_{\sigma\sigma}^{<} & = & g^{<}+g^{r}UP^{r}g_{2}^{<}\left\langle n_{\bar{\sigma}}\right\rangle +g^{<}UP^{a}g_{2}^{a}\left(\left\langle n_{\bar{\sigma}}\right\rangle -\Sigma_{1}^{a}g^{a}\right)\nonumber \\
 &  & -g^{r}UP^{r}g_{2}^{r}\left(\Sigma_{1}^{r}g^{<}+\Sigma_{1}^{<}g^{a}\right)-g^{r}UP^{r}g_{2}^{<}\Sigma_{1}^{a}g^{a}\nonumber \\
 &  & -g^{r}UP^{r}g_{2}^{r}\Sigma_{1}^{r}g^{<}UP^{a}g_{2}^{a}\left(\left\langle n_{\bar{\sigma}}\right\rangle +\Sigma_{1}^{a}g^{a}\right)\\
 &  & -g^{r}UP^{r}g_{2}^{r}\Sigma_{1}^{<}g^{a}UP^{a}g_{2}^{a}\left(\left\langle n_{\bar{\sigma}}\right\rangle +\Sigma_{1}^{a}g^{a}\right)\nonumber \\
 &  & -g^{r}UP^{r}g_{2}^{<}\Sigma_{1}^{a}g^{a}UP^{a}g_{2}^{a}\left(\left\langle n_{\bar{\sigma}}\right\rangle +\Sigma_{1}^{a}g^{a}\right),\nonumber 
\end{eqnarray}
where $P^{r,a}=\frac{1}{1+g_{2}^{r,a}\Sigma_{1}^{r,a}g^{r,a}U}$,
$g^{<}=g^{r}\Sigma_{0}^{<}g^{a}$, and $g_{2}^{<}=g_{2}^{r}\Sigma_{4}^{<}g_{2}^{a}$
. Applying the principle of reductio ad absurdum we assume $G_{\sigma\sigma}^{<}$
is imaginary. Since it must hold for any real value of $\left\langle n_{\bar{\sigma}}\right\rangle $
between $0$ and $1$, we argue that the term
\begin{eqnarray}
A_{1} & = & g^{r}UP^{r}g_{2}^{<}\left\langle n_{\bar{\sigma}}\right\rangle +g^{<}UP^{a}g_{2}^{a}\left\langle n_{\bar{\sigma}}\right\rangle \nonumber \\
 &  & -g^{r}UP^{r}g_{2}^{r}\Sigma_{1}^{r}g^{<}UP^{a}g_{2}^{a}\left\langle n_{\bar{\sigma}}\right\rangle \nonumber \\
 &  & -g^{r}UP^{r}g_{2}^{r}\Sigma_{1}^{<}g^{a}UP^{a}g_{2}^{a}\left\langle n_{\bar{\sigma}}\right\rangle \\
 &  & -g^{r}UP^{r}g_{2}^{<}\Sigma_{1}^{a}g^{a}UP^{a}g_{2}^{a}\left\langle n_{\bar{\sigma}}\right\rangle ,\nonumber 
\end{eqnarray}
must be imaginary. Moreover, Since $A_{1}$ must be imaginary for
any value of $U$ the term 
\begin{eqnarray}
A_{2} & = & g^{r}UP^{r}g_{2}^{<}\left\langle n_{\bar{\sigma}}\right\rangle +g^{<}UP^{a}g_{2}^{a}\left\langle n_{\bar{\sigma}}\right\rangle ,
\end{eqnarray}
should be imaginary as well. Using the fact that $U$ and $\left\langle n_{\bar{\sigma}}\right\rangle $
are real quantities and by definition $g_{2}^{<}$ and $g^{<}$ are
imaginary, for $A_{2}$ to be imaginary the following must hold:
\begin{equation}
Im\left(g^{r}P^{r}\right)g_{2}^{<}=-Im\left(P^{a}g_{2}^{a}\right)g^{<},\label{eq:demand1}
\end{equation}
or in other words, we demand that $\Re\left(A_{2}\right)=0.$ One
can then show (see online supporting material for more information)
that, in fact, the equality in equation (\ref{eq:demand1}) does not
hold, namely, $G_{\sigma\sigma}^{<}\left(\omega\right)$ is not an
imaginary function and the relation $G_{\sigma\sigma}^{<}\left(\omega\right)=-\left(G_{\sigma\sigma}^{<}\left(\omega\right)\right)^{*}$
is not satisfied. In turn, this implies that $\left\langle n_{\sigma}\right\rangle $
(the occupation number) is a complex number, which of course is not
physical. Following the same derivation one can show that $G_{\sigma\sigma}^{>}\left(\omega\right)$
is not an imaginary function either. All the other relations given
in equation (\ref{eq:relations}) are fulfilled. 

If one is only interested in the Coulomb blockade regime, it is not
necessary to go to the level of approximation presented here (which
is essential to obtain the Kondo effect). For the Coulomb blockade
regime one can turn to the approximation presented in Refs.~\onlinecite{Song2007},
where on top of the approximations described above we also neglect
the simultaneous hopping of electron pairs to and from the system.
This approximation does not violet the symmetry relations of the single
particle GF (see online supporting information for further discussion),
but as pointed above, it does not reproduce the Kondo peaks at low
temperatures.

\subsection{The double Anderson model\label{sub:The-double-Anderson}}

For the double Anderson model we follow the derivation given in Ref.~\onlinecite{Joshi2000},
and define the following contour ordered GF
\begin{equation}
G_{\alpha\beta}^{\sigma\sigma}\left(t,t'\right)=-\frac{i}{\hbar}\left\langle T_{C}d_{\alpha\sigma}\left(t\right)d_{\beta\sigma}^{\dagger}\left(t'\right)\right\rangle ,
\end{equation}
\begin{equation}
\mathbb{\mathbb{G}_{\alpha\beta\gamma}^{\tau\sigma\sigma}}\left(t,t'\right)=-\frac{i}{\hbar}\left\langle T_{C}n_{\alpha\tau}\left(t\right)d_{\beta\sigma}\left(t\right)d_{\gamma\sigma}^{\dagger}\left(t'\right)\right\rangle ,
\end{equation}
where $\tau=\sigma,\bar{\sigma}$ . The approximations used in Ref.~\onlinecite{Joshi2000}
are: (a) neglect the simultaneous hopping of electron pairs to and
from the system, (b) assume that $F_{2}\left(t,t'\right)=-\frac{i}{\hbar}\left\langle T_{C}c_{ki\sigma}\left(t\right)n\left(t\right)d_{\beta\sigma}^{\dagger}\left(t'\right)\right\rangle \approx-\frac{i}{\hbar}\sum_{\gamma=\alpha,\beta}t_{k,\gamma}^{\sigma}\int\mbox{d}t_{1}g_{k}\left(t,t_{1}\right)\left\langle T_{C}d_{\gamma\sigma}\left(t_{1}\right)n\left(t_{1}\right)d_{\beta\sigma}^{\dagger}\left(t'\right)\right\rangle $
where $n\left(t\right)$ is the number operator of one of the electrons
of the system, and $\left(i\hbar\frac{\partial}{\partial t}-\varepsilon_{k\sigma}\right)g_{k}\left(t,t_{1}\right)=\delta\left(t-t_{1}\right)$,
and (c) higher order GFs of the form $-\frac{i}{\hbar}\left\langle T_{c}\left[n_{\gamma,\sigma}\left(t\right)n_{\delta,\tau}\left(t\right)d_{\alpha,\sigma}\left(t\right)d_{\beta,\sigma}^{\dagger}\left(0\right)\right]\right\rangle $
are decoupled to $-\frac{i}{\hbar}\left\langle n_{\gamma,\sigma}\left(t\right)\right\rangle \left\langle T_{c}n_{\delta,\tau}\left(t\right)d_{\alpha,\sigma}\left(t\right)d_{\beta,\sigma}^{\dagger}\left(0\right)\right\rangle -\frac{i}{\hbar}\left\langle n_{\delta,\sigma}\left(t\right)\right\rangle \left\langle T_{c}n_{\gamma,\sigma}\left(t\right)d_{\alpha,\sigma}\left(t\right)d_{\beta,\sigma}^{\dagger}\left(0\right)\right\rangle $
. These approximations lead to the following results
\begin{eqnarray}
G_{\alpha\beta}^{\sigma\sigma}\left(\omega\right) & = & \left(\hbar\omega-\varepsilon_{\alpha,\sigma}-\Sigma_{0}\left(\omega\right)\right)^{-1}\times\left(\delta_{\alpha\beta}^{\sigma\sigma}\right.\nonumber \\
 &  & +h_{\alpha\beta}^{\sigma}G_{\beta\beta}^{\sigma\sigma}\left(\omega\right)+U_{\alpha}\mathbb{G}_{\alpha\alpha\beta}^{\bar{\sigma}\sigma\sigma}\left(\omega\right)\label{eq:Single part}\\
 &  & \left.+V_{\alpha\beta}^{\sigma\bar{\sigma}}\mathbb{G}_{\beta\alpha\beta}^{\bar{\sigma}\sigma\sigma}\left(\omega\right)+V_{\alpha\beta}^{\sigma\sigma}\mathbb{G}_{\beta\alpha\beta}^{\sigma\sigma\sigma}\left(\omega\right)\right),\nonumber 
\end{eqnarray}
\begin{widetext}
\begin{eqnarray}
\mathbb{G}_{\alpha\alpha\beta}^{\bar{\sigma}\sigma\sigma}\left(\omega\right) & = & \left(\hbar\omega-\varepsilon_{\alpha\sigma}-U_{\alpha}-V_{\alpha\beta}^{\sigma\sigma}\left\langle n_{\beta\sigma}\right\rangle -V_{\alpha\beta}^{\sigma\bar{\sigma}}\left\langle n_{\beta\bar{\sigma}}\right\rangle -\Sigma_{0}\left(\omega\right)\right)^{-1}\nonumber \\
 &  & \times\left[h_{\alpha\beta}^{\sigma}\mathbb{G}_{\alpha\beta\beta}^{\bar{\sigma}\sigma\sigma}\left(\omega\right)+\left\langle n_{\alpha\bar{\sigma}}\right\rangle \left(V_{\alpha\beta}^{\sigma\sigma}\mathbb{G}_{\beta\alpha\beta}^{\sigma\sigma\sigma}\left(\omega\right)+V_{\alpha\beta}^{\sigma\bar{\sigma}}\mathbb{G}_{\beta\alpha\beta}^{\bar{\sigma}\sigma\sigma}\left(\omega\right)\right)\right],\nonumber \\
\mathbb{G}_{\alpha\beta\beta}^{\bar{\sigma}\sigma\sigma}\left(\omega\right) & = & \left(\hbar\omega-\varepsilon_{\beta\sigma}-U_{\beta}\left\langle n_{\beta\bar{\sigma}}\right\rangle -V_{\beta\alpha}^{\sigma\sigma}\left\langle n_{\alpha\sigma}\right\rangle -V_{\beta\alpha}^{\sigma\bar{\sigma}}-\Sigma_{0}\left(\omega\right)\right)^{-1}\nonumber \\
 &  & \times\left[\left\langle n_{\alpha\bar{\sigma}}\right\rangle +h_{\beta\alpha}^{\sigma}\mathbb{G}_{\alpha\alpha\beta}^{\bar{\sigma}\sigma\sigma}\left(\omega\right)+\left\langle n_{\alpha\bar{\sigma}}\right\rangle \left(U_{\beta}\mathbb{G}_{\beta\beta\beta}^{\bar{\sigma}\sigma\sigma}\left(\omega\right)+V_{\beta\alpha}^{\sigma\sigma}\mathbb{G}_{\alpha\beta\beta}^{\sigma\sigma\sigma}\left(\omega\right)\right)\right],\nonumber \\
\mathbb{G}_{\alpha\beta\beta}^{\sigma\sigma\sigma}\left(\omega\right) & = & \left(\hbar\omega-\varepsilon_{\beta\sigma}-U_{\beta}\left\langle n_{\beta\bar{\sigma}}\right\rangle -V_{\beta\alpha}^{\sigma\bar{\sigma}}\left\langle n_{\alpha\bar{\sigma}}\right\rangle -V_{\beta\alpha}^{\sigma\sigma}-\Sigma_{0}\left(\omega\right)\right)^{-1}\nonumber \\
 &  & \times\left[\left\langle n_{\alpha\sigma}\right\rangle +h_{\beta\alpha}^{\sigma}\mathbb{G}_{\beta\alpha\beta}^{\sigma\sigma\sigma}\left(\omega\right)+\left\langle n_{\alpha\sigma}\right\rangle \left(U_{\beta}\mathbb{G}_{\beta\beta\beta}^{\bar{\sigma}\sigma\sigma}\left(\omega\right)+V_{\beta\alpha}^{\sigma\bar{\sigma}}\mathbb{G}_{\alpha\beta\beta}^{\bar{\sigma}\sigma\sigma}\left(\omega\right)\right)\right],\nonumber \\
\mathbb{G}_{\beta\alpha\beta}^{\bar{\sigma}\sigma\sigma}\left(\omega\right) & = & \left(\hbar\omega-\varepsilon_{\alpha\sigma}-U_{\alpha}\left\langle n_{\alpha\bar{\sigma}}\right\rangle -V_{\alpha\beta}^{\sigma\sigma}\left\langle n_{\beta\sigma}\right\rangle -V_{\alpha\beta}^{\sigma\sigma}-\Sigma_{0}\left(\omega\right)\right)^{-1}\\
 &  & \times\left[h_{\alpha\beta}^{\sigma}\mathbb{G}_{\beta\beta\beta}^{\bar{\sigma}\sigma\sigma}\left(\omega\right)+\left\langle n_{\beta\bar{\sigma}}\right\rangle \left(U_{\alpha}\mathbb{G}_{\alpha\alpha\beta}^{\bar{\sigma}\sigma\sigma}\left(\omega\right)+V_{\alpha\beta}^{\sigma,\sigma}\mathbb{G}_{\beta\alpha\beta}^{\sigma\sigma\sigma}\left(\omega\right)\right)\right],\nonumber \\
\mathbb{G}_{\beta\alpha\beta}^{\sigma\sigma\sigma}\left(\omega\right) & = & \left(\hbar\omega-\varepsilon_{\alpha\sigma}-U_{\alpha}\left\langle n_{\alpha\bar{\sigma}}\right\rangle -V_{\alpha\beta}^{\sigma\bar{\sigma}}\left\langle n_{\beta\bar{\sigma}}\right\rangle -V_{\alpha\beta}^{\sigma\sigma}-\Sigma_{0}\left(\omega\right)\right)^{-1}\nonumber \\
 &  & \times\left[-\left\langle d_{\beta\sigma}^{\dagger}d_{\alpha,\sigma}\right\rangle +h_{\alpha\beta}^{\sigma}\mathbb{G}_{\alpha\beta\beta}^{\sigma\sigma\sigma}\left(\omega\right)+\left\langle n_{\beta,\sigma}\right\rangle \left(U_{\alpha}\mathbb{G}_{\alpha\alpha\beta}^{\bar{\sigma}\sigma\sigma}\left(\omega\right)+V_{\alpha,\beta}^{\sigma,\bar{\sigma}}\mathbb{G}_{\beta\alpha\beta}^{\bar{\sigma}\sigma\sigma}\left(\omega\right)\right)\right],\nonumber \\
\mathbb{G}_{\beta\beta\beta}^{\bar{\sigma}\sigma\sigma}\left(\omega\right) & = & \left(\hbar\omega-\varepsilon_{\beta\sigma}-U_{\beta}-V_{\beta\alpha}^{\sigma\bar{\sigma}}\left\langle n_{\alpha\bar{\sigma}}\right\rangle -V_{\beta\alpha}^{\sigma\sigma}\left\langle n_{\alpha\sigma}\right\rangle -\Sigma_{0}\left(\omega\right)\right)^{-1}\nonumber \\
 &  & \times\left[\left\langle n_{\beta\bar{\sigma}}\right\rangle +h_{\beta\alpha}^{\sigma}\mathbb{G}_{\beta\alpha\beta}^{\bar{\sigma}\sigma\sigma}\left(\omega\right)+\left\langle n_{\beta\bar{\sigma}}\right\rangle \left(V_{\beta\alpha}^{\sigma,\sigma}\mathbb{G}_{\alpha\beta\beta}^{\sigma\sigma\sigma}\left(\omega\right)+V_{\beta\alpha}^{\sigma,\bar{\sigma}}\mathbb{G}_{\alpha\beta\beta}^{\bar{\sigma}\sigma\sigma}\left(\omega\right)\right)\right],\nonumber 
\end{eqnarray}

\end{widetext}~\\
We now show that given this set of equations, the symmetry relation
$\left(G_{\alpha\beta}^{\sigma\sigma}\left(\omega\right)\right)^{r}=\left(\left(G_{\beta\alpha}^{\sigma\sigma}\left(\omega\right)\right)^{a}\right)^{*}$
is not satisfied. By applying the Langreth rules we can find the retarded
and advanced projections of the single particle GF (equation (\ref{eq:Single part})).
For simplicity we derived them for the case where $V_{ij}^{\sigma\tau}=0$.
Define 
\begin{eqnarray}
\left(g_{i}\right)^{r,a} & = & \frac{1}{\hbar\omega-\varepsilon_{i,\sigma}-\Sigma_{0}^{r,a}},\\
\left(g_{ii}^{\bar{\sigma}\sigma}\right)^{r,a} & = & \frac{1}{\hbar\omega-\varepsilon_{i,\sigma}-U_{i}-\Sigma_{0}^{r,a}},\\
\left(g_{ij}^{\bar{\sigma}\sigma}\right)^{r,a} & = & \frac{1}{\hbar\omega-\varepsilon_{j,\sigma}-U_{j}\left\langle n_{j,\bar{\sigma}}\right\rangle -\Sigma_{0}^{r,a}}.
\end{eqnarray}
Given these definitions, the retarded and advanced GFs are given by:\\

\begin{widetext}
\begin{eqnarray}
\left(G_{\alpha\beta}^{\sigma\sigma}\right)^{r} & = & \left(I-\left(g_{\alpha}\right)^{r}h_{\alpha\beta}^{\sigma}\left(g_{\beta}\right)^{r}h_{\beta,\alpha}^{\sigma}\right)^{-1}\label{eq:gr}\\
 &  & \times\left(\left(g_{\alpha}\right)^{r}h_{\alpha\beta}^{\sigma}\left(g_{\beta}\right)^{r}+\left(g_{\alpha}\right)^{r}h_{\alpha\beta}^{\sigma}\left(g_{\beta}\right)^{r}U_{\beta}\left(\mathbb{G}_{\beta\beta\beta}^{\bar{\sigma}\sigma\sigma}\right)^{r}+\left(g_{\alpha}\right)^{r}U_{\alpha}\left(\mathbb{G}_{\alpha\alpha\beta}^{\bar{\sigma}\sigma\sigma}\right)^{r}\right),\nonumber 
\end{eqnarray}
\begin{eqnarray}
\left(\mathbb{G}_{\alpha\alpha\beta}^{\bar{\sigma}\sigma\sigma}\right)^{r} & = & \left(1-\left(g_{\alpha\alpha}^{\bar{\sigma}\sigma}\right)^{r}h_{\alpha\beta}^{\sigma}\left(g_{\alpha\beta}^{\bar{\sigma}\sigma}\right)^{r}h_{\beta\alpha}^{\sigma}-\left(g_{\alpha\alpha}^{\bar{\sigma}\sigma}\right)^{r}h_{\alpha\beta}^{\sigma}\left(g_{\alpha\beta}^{\bar{\sigma}\sigma}\right)^{r}\left\langle n_{\alpha\bar{\sigma}}\right\rangle U_{\beta}\right.\nonumber \\
 &  & \times\left.\left(1-\left(g_{\beta\beta}^{\bar{\sigma}\sigma}\right)^{r}h_{\beta\alpha}^{\sigma}\left(g_{\beta\alpha}^{\bar{\sigma}\sigma}\right)^{r}h_{\alpha\beta}^{\sigma}\right)^{-1}\left(g_{\beta\beta}^{\bar{\sigma}\sigma}\right)^{r}h_{\beta\alpha}^{\sigma}\left(g_{\beta\alpha}^{\bar{\sigma}\sigma}\right)^{r}\left\langle n_{\beta\bar{\sigma}}\right\rangle U_{\alpha}\right)^{-1}\\
 &  & \times\left(\left(g_{\alpha\alpha}^{\bar{\sigma}\sigma}\right)^{r}h_{\alpha\beta}^{\sigma}\left(g_{\alpha\beta}^{\bar{\sigma}\sigma}\right)^{r}\left\langle n_{\alpha\bar{\sigma}}\right\rangle +\left(g_{\alpha\alpha}^{\bar{\sigma}\sigma}\right)^{r}h_{\alpha\beta}^{\sigma}\left(g_{\alpha\beta}^{\bar{\sigma}\sigma}\right)^{r}\left\langle n_{\alpha\bar{\sigma}}\right\rangle U_{\beta}\right.\nonumber \\
 &  & \times\left.\left(1-\left(g_{\beta\beta}^{\bar{\sigma}\sigma}\right)^{r}h_{\beta\alpha}^{\sigma}\left(g_{\beta\alpha}^{\bar{\sigma}\sigma}\right)^{r}h_{\alpha\beta}^{\sigma}\right)^{-1}\left(g_{\beta\beta}^{\bar{\sigma}\sigma}\right)^{r}\left\langle n_{\beta\bar{\sigma}}\right\rangle \right),\nonumber 
\end{eqnarray}
\begin{eqnarray}
\left(\mathbb{G}_{\beta\beta\beta}^{\bar{\sigma}\sigma\sigma}\right)^{r} & = & \left(1-\left(g_{\beta\beta}^{\bar{\sigma}\sigma}\right)^{r}h_{\beta\alpha}^{\sigma}\left(g_{\beta\alpha}^{\bar{\sigma}\sigma}\right)^{r}h_{\alpha\beta}^{\sigma}-\left(g_{\beta\beta}^{\bar{\sigma}\sigma}\right)^{r}h_{\beta\alpha}^{\sigma}\left(g_{\beta\alpha}^{\bar{\sigma}\sigma}\right)^{r}\left\langle n_{\beta\bar{\sigma}}\right\rangle U_{\alpha}\right.\nonumber \\
 &  & \times\left.\left(1-\left(g_{\alpha\alpha}^{\bar{\sigma}\sigma}\right)^{r}h_{\alpha\beta}^{\sigma}\left(g_{\alpha\beta}^{\bar{\sigma}\sigma}\right)^{r}h_{\beta\alpha}^{\sigma}\right)^{-1}\left(g_{\alpha\alpha}^{\bar{\sigma}\sigma}\right)^{r}h_{\alpha\beta}^{\sigma}\left(g_{\alpha\beta}^{\bar{\sigma}\sigma}\right)^{r}\left\langle n_{\alpha\bar{\sigma}}\right\rangle U_{\beta}\right)^{-1}\\
 &  & \times\left(\left(g_{\beta\beta}^{\bar{\sigma}\sigma}\right)^{r}\left\langle n_{\beta\bar{\sigma}}\right\rangle +\left(g_{\beta\beta}^{\bar{\sigma}\sigma}\right)^{r}h_{\beta\alpha}^{\sigma}\left(g_{\beta\alpha}^{\bar{\sigma}\sigma}\right)^{r}\left\langle n_{\beta\bar{\sigma}}\right\rangle U_{\alpha}\right.\nonumber \\
 &  & \times\left.\left(1-\left(g_{\alpha\alpha}^{\bar{\sigma}\sigma}\right)^{r}h_{\alpha\beta}^{\sigma}\left(g_{\alpha\beta}^{\bar{\sigma}\sigma}\right)^{r}h_{\beta\alpha}^{\sigma}\right)^{-1}\left(g_{\alpha\alpha}^{\bar{\sigma}\sigma}\right)^{r}h_{\alpha\beta}^{\sigma}\left(g_{\alpha\beta}^{\bar{\sigma}\sigma}\right)^{r}\left\langle n_{\alpha\bar{\sigma}}\right\rangle \right),\nonumber 
\end{eqnarray}
\begin{eqnarray}
\left(G_{\beta\alpha}^{\sigma\sigma}\right)^{a} & = & \left(I-\left(g_{\beta}\right)^{a}h_{\beta\alpha}^{\sigma}\left(g_{\alpha}\right)^{a}h_{\alpha\beta}^{\sigma}\right)^{-1}\label{eq:ga}\\
 &  & \times\left(\left(g_{\beta}\right)^{a}h_{\beta\alpha}^{\sigma}\left(g_{\alpha}\right)^{a}+\left(g_{\beta}\right)^{a}h_{\beta\alpha}^{\sigma}\left(g_{\alpha}\right)^{a}U_{\alpha}\left(\mathbb{G}_{\alpha\alpha\alpha}^{\bar{\sigma}\sigma\sigma}\right)^{a}+\left(g_{\beta}\right)^{a}U_{\beta}\left(\mathbb{G}_{\beta\beta\alpha}^{\bar{\sigma}\sigma\sigma}\right)^{a}\right),\nonumber 
\end{eqnarray}
\begin{eqnarray}
\left(\mathbb{G}_{\beta\beta\alpha}^{\bar{\sigma}\sigma\sigma}\right)^{a} & = & \left(1-\left(g_{\beta\beta}^{\bar{\sigma}\sigma}\right)^{a}h_{\beta\alpha}^{\sigma}\left(g_{\beta\alpha}^{\bar{\sigma}\sigma}\right)^{a}h_{\alpha\beta}^{\sigma}-\left(g_{\beta\beta}^{\bar{\sigma}\sigma}\right)^{a}h_{\beta\alpha}^{\sigma}\left(g_{\beta\alpha}^{\bar{\sigma}\sigma}\right)^{a}\left\langle n_{\beta\bar{\sigma}}\right\rangle U_{\alpha}\right.\nonumber \\
 &  & \times\left.\left(1-\left(g_{\alpha\alpha}^{\bar{\sigma}\sigma}\right)^{a}h_{\alpha\beta}^{\sigma}\left(g_{\alpha\beta}^{\bar{\sigma}\sigma}\right)^{a}h_{\beta\alpha}^{\sigma}\right)^{-1}\left(g_{\alpha\alpha}^{\bar{\sigma}\sigma}\right)^{a}h_{\alpha\beta}^{\sigma}\left(g_{\alpha\beta}^{\bar{\sigma}\sigma}\right)^{a}\left\langle n_{\alpha\bar{\sigma}}\right\rangle U_{\beta}\right)^{-1}\\
 &  & \times\left(\left(g_{\beta\beta}^{\bar{\sigma}\sigma}\right)^{a}h_{\beta\alpha}^{\sigma}\left(g_{\beta\alpha}^{\bar{\sigma}\sigma}\right)^{a}\left\langle n_{\beta\bar{\sigma}}\right\rangle +\left(g_{\beta\beta}^{\bar{\sigma}\sigma}\right)^{a}h_{\beta\alpha}^{\sigma}\left(g_{\beta\alpha}^{\bar{\sigma}\sigma}\right)^{a}\left\langle n_{\beta\bar{\sigma}}\right\rangle U_{\alpha}\right.\nonumber \\
 &  & \times\left.\left(1-\left(g_{\alpha\alpha}^{\bar{\sigma}\sigma}\right)^{a}h_{\alpha\beta}^{\sigma}\left(g_{\alpha\beta}^{\bar{\sigma}\sigma}\right)^{a}h_{\beta\alpha}^{\sigma}\right)^{-1}\left(g_{\alpha\alpha}^{\bar{\sigma}\sigma}\right)^{a}\left\langle n_{\alpha\bar{\sigma}}\right\rangle \right),\nonumber 
\end{eqnarray}
\begin{eqnarray}
\left(\mathbb{G}_{\alpha\alpha\alpha}^{\bar{\sigma}\sigma\sigma}\right)^{a} & = & \left(1-\left(g_{\alpha\alpha}^{\bar{\sigma}\sigma}\right)^{a}h_{\alpha\beta}^{\sigma}\left(g_{\alpha\beta}^{\bar{\sigma}\sigma}\right)^{a}h_{\beta\alpha}^{\sigma}-\left(g_{\alpha\alpha}^{\bar{\sigma}\sigma}\right)^{a}h_{\alpha\beta}^{\sigma}\left(g_{\alpha\beta}^{\bar{\sigma}\sigma}\right)^{a}\left\langle n_{\alpha\bar{\sigma}}\right\rangle U_{\beta}\right.\nonumber \\
 &  & \times\left.\left(1-\left(g_{\beta\beta}^{\bar{\sigma}\sigma}\right)^{a}h_{\beta\alpha}^{\sigma}\left(g_{\beta\alpha}^{\bar{\sigma}\sigma}\right)^{a}h_{\alpha\beta}^{\sigma}\right)^{-1}\left(g_{\beta\beta}^{\bar{\sigma}\sigma}\right)^{a}h_{\beta\alpha}^{\sigma}\left(g_{\beta\alpha}^{\bar{\sigma}\sigma}\right)^{a}\left\langle n_{\beta\bar{\sigma}}\right\rangle U_{\alpha}\right)^{-1}\\
 &  & \times\left(\left(g_{\alpha\alpha}^{\bar{\sigma}\sigma}\right)^{a}\left\langle n_{\alpha\bar{\sigma}}\right\rangle +\left(g_{\alpha\alpha}^{\bar{\sigma}\sigma}\right)^{a}h_{\alpha\beta}^{\sigma}\left(g_{\alpha\beta}^{\bar{\sigma}\sigma}\right)^{a}\left\langle n_{\alpha\bar{\sigma}}\right\rangle U_{\beta}\right.\nonumber \\
 &  & \times\left.\left(1-\left(g_{\beta\beta}^{\bar{\sigma}\sigma}\right)^{a}h_{\beta\alpha}^{\sigma}\left(g_{\beta\alpha}^{\bar{\sigma}\sigma}\right)^{a}h_{\alpha\beta}^{\sigma}\right)^{-1}\left(g_{\beta\beta}^{\bar{\sigma}\sigma}\right)^{a}h_{\beta\alpha}^{\sigma}\left(g_{\beta\alpha}^{\bar{\sigma}\sigma}\right)^{a}\left\langle n_{\beta\bar{\sigma}}\right\rangle \right).\nonumber 
\end{eqnarray}
\end{widetext}Substituting the equations for $\left(\mathbb{G}_{\alpha\alpha\beta}^{\bar{\sigma}\sigma\sigma}\left(\omega\right)\right)^{r}$,
$\left(\mathbb{G}_{\beta\beta\beta}^{\bar{\sigma}\sigma\sigma}\left(\omega\right)\right)^{r}$,
$\left(\mathbb{G}_{\beta\beta\alpha}^{\bar{\sigma}\sigma\sigma}\left(\omega\right)\right)^{a}$
and $\left(\mathbb{G}_{\alpha\alpha\alpha}^{\bar{\sigma}\sigma\sigma}\left(\omega\right)\right)^{a}$
into equations (\ref{eq:gr}) and (\ref{eq:ga}), respectively, and
comparing the resulting expressions we find that $\left(G_{\alpha\beta}^{\sigma\sigma}\left(\omega\right)\right)^{r}\neq\left(\left(G_{\beta\alpha}^{\sigma\sigma}\left(\omega\right)\right)^{a}\right)^{*}$
(see online supporting material for more details). Moreover, we find
that none of the symmetry relations in equation (\ref{eq:relations})
hold. In the following section we propose a symmetrization scheme
that restores all the symmetries of the single particle GF.

\section{Symmetry restoration\label{sec:Symmetry-restoration}}

\subsection{Guidelines to restore symmetry \label{sub:Symmetry-restoration}}

The customary route to calculate the NEGF is as follows: (a) calculate
the retarded GF and use it to obtain the advanced GF (by demanding
$G_{\alpha\beta}^{a}\left(\omega\right)=\left(G_{\beta\alpha}^{r}\left(\omega\right)\right)^{*}$).
(b) Calculate the lesser/greater GF and symmetrize the lesser/greater
to fulfill the quantum Onsager relations~\citealp{Bulka2004}, hence
obeying $G_{\alpha\beta}^{<,>}\left(\omega\right)=-\left(G_{\beta\alpha}^{<,>}\left(\omega\right)\right)^{*}$.
In most applications of NEGF the advanced GF is not directly calculated
and thus, the symmetry breakage does not always stand out. In fact,
this common procedure restores the relation between the advanced and
retarded GF and between the lesser/greater and their complex conjugate,
but does not necessarily restore the relation $G_{\alpha\beta}^{r}\left(\omega\right)-G_{\alpha\beta}^{a}\left(\omega\right)=G_{\alpha\beta}^{>}\left(\omega\right)-G_{\alpha\beta}^{<}\left(\omega\right)$.
It can be shown that violation of the latter leads to violation of
the fluctuation dissipation relation, $\mathbf{G}^{<}=-f_{eq}\left(\varepsilon-\mu_{eq}\right)\left(\mathbf{G}^{r}-\mathbf{G}^{a}\right)$,
at equilibrium. This oversimplified procedure can result in different
values for the currents depending on how it is calculated, cf. equation
(\ref{eq:I1}) or equation (\ref{eq:I2}). It may also lead to finite
currents at zero-bias voltage (see Sec. \ref{sub:Symmetrization-of-the-DA}
for more), which is physically incorrect.

In order to restore the symmetry relations that are imposed by the
definitions of the GF (cf. equation \ref{eq:relations}), we suggest
the following procedure:
\begin{enumerate}
\item Calculate the retarded/advanced GFs matrices $\left(\mathbf{G}^{r}/\mathbf{G}^{a}\right)$
separately and use them to define ``new'' retarded/advanced GFs
matrices $\tilde{\mathbf{G}}^{r}=\frac{1}{2}\left(\mathbf{G}^{r}+\left(\mathbf{G}^{a}\right)^{\dagger}\right)$
and $\tilde{\mathbf{G}}^{a}=\frac{1}{2}\left(\mathbf{G}^{a}+\left(\mathbf{G}^{r}\right)^{\dagger}\right)=\left(\tilde{\mathbf{G}}^{r}\right)^{\dagger}$. 
\item Use the ``new'' retarded/advanced GFs matrices $\left(\tilde{\mathbf{G}}^{r}/\tilde{\mathbf{G}}^{a}\right)$
to calculate the lesser/greater GFs matrices $\left(\mathbf{G}^{<}/\mathbf{G}^{>}\right)$.
Again, use them to define ``new'' lesser/greater GFs matrices $\tilde{\mathbf{G}}^{<,>}=\frac{1}{2}\left(\mathbf{G}^{<,>}-\left(\mathbf{G}^{<,>}\right)^{\dagger}\right)$. 
\item Calculate the two anti-Hermitian matrices $\mathbf{A}=\tilde{\mathbf{G}}^{>}-\tilde{\mathbf{G}}^{<}$
and $\mathbf{B}=\tilde{\mathbf{G}}^{r}-\tilde{\mathbf{G}}^{a}$. Define
the difference anti-Hermitian matrix $\mathbf{C}=\mathbf{A}-\mathbf{B}$,
and redefine the retarded and advanced GFs $\bar{\mathbf{G}}^{r}=\tilde{\mathbf{G}}^{r}+\frac{\mathbf{C}}{2}$,
and $\bar{\mathbf{G}}^{a}=\tilde{\mathbf{G}}^{a}-\frac{\mathbf{C}}{2}$.
\end{enumerate}
The resulting GFs ($\bar{\mathbf{G}}^{r},\,\bar{\mathbf{G}}^{a},\,\tilde{\mathbf{G}}^{<}$
and $\tilde{\mathbf{G}}^{>}$) obey all symmetry relations of equation
(\ref{eq:relations}) by construction. Note that if the original GFs
obeyed the symmetry relations to begin with, our symmetrization procedure
will not alter them in any way. 

We now turn to perform detailed calculations for both the Anderson
and double Anderson models. For the Anderson model, we use the closure
described in Sec. \ref{sub:The-Anderson-model} while for the double
Anderson model we use the closure described in Sec. \ref{sub:The-double-Anderson}.
The resulting EOMs were solved self-consistently in Fourier space
with a frequency discretization of $N_{\omega}=2^{14}-2^{16}$ depending
on the model parameters. Typically, $<15$ self-consistent iterations
were needed to converge the results. Convergence was declared when
the population values at subsequent iteration steps did not change
within a predefined tolerance value chosen as $10^{-6}$. For each
set of calculations we have applied the above symmetrization scheme
and compared the results to those obtained without restoring symmetry,
as detailed for each model.

\subsection{Anderson impurity model\label{sub:Symmetrization-of-the-A }}

\begin{figure}[h]
\centering{}\includegraphics[width=8cm]{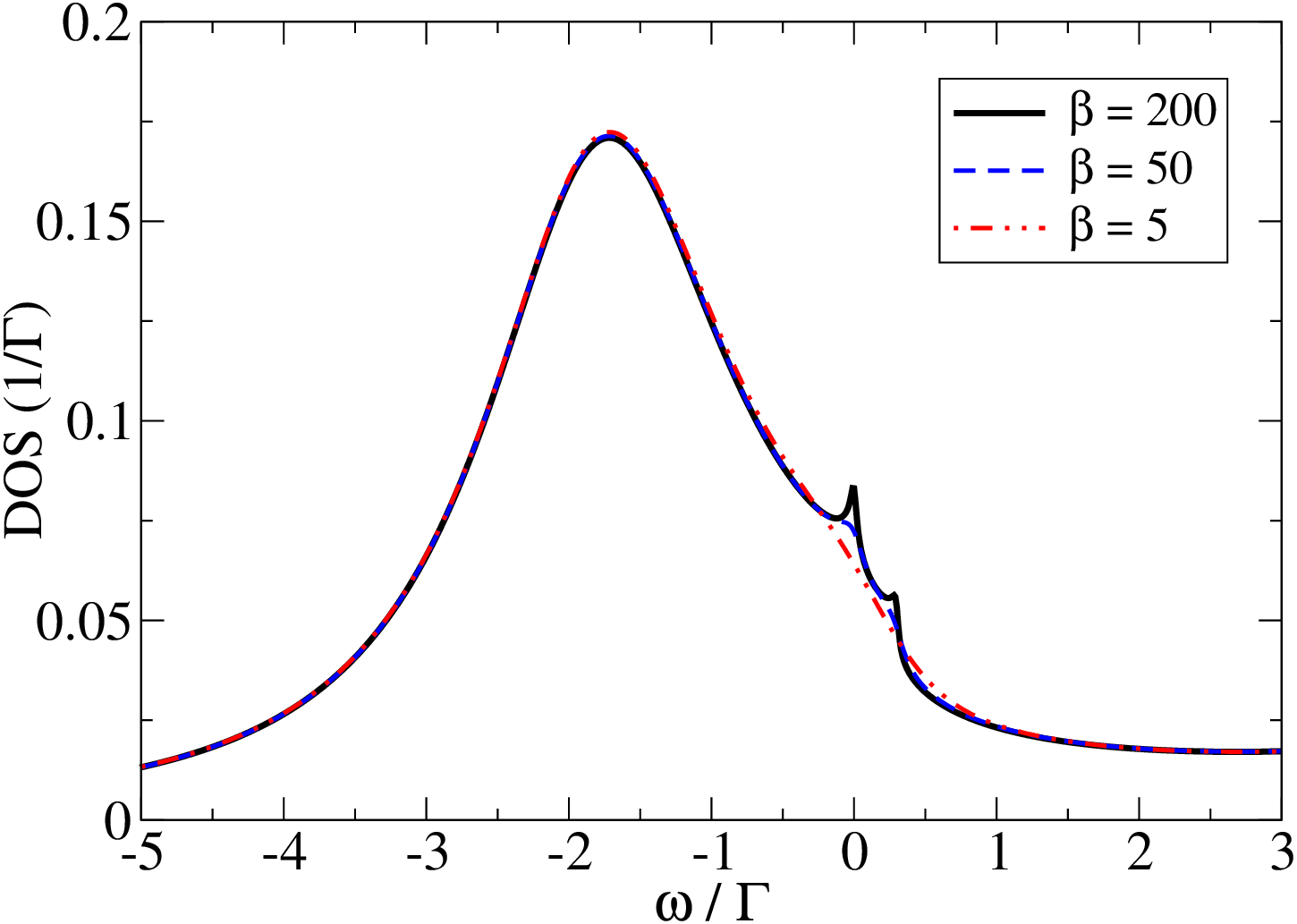}\\ \ \\ \ \\ \ \\ 
\includegraphics[width=8cm]{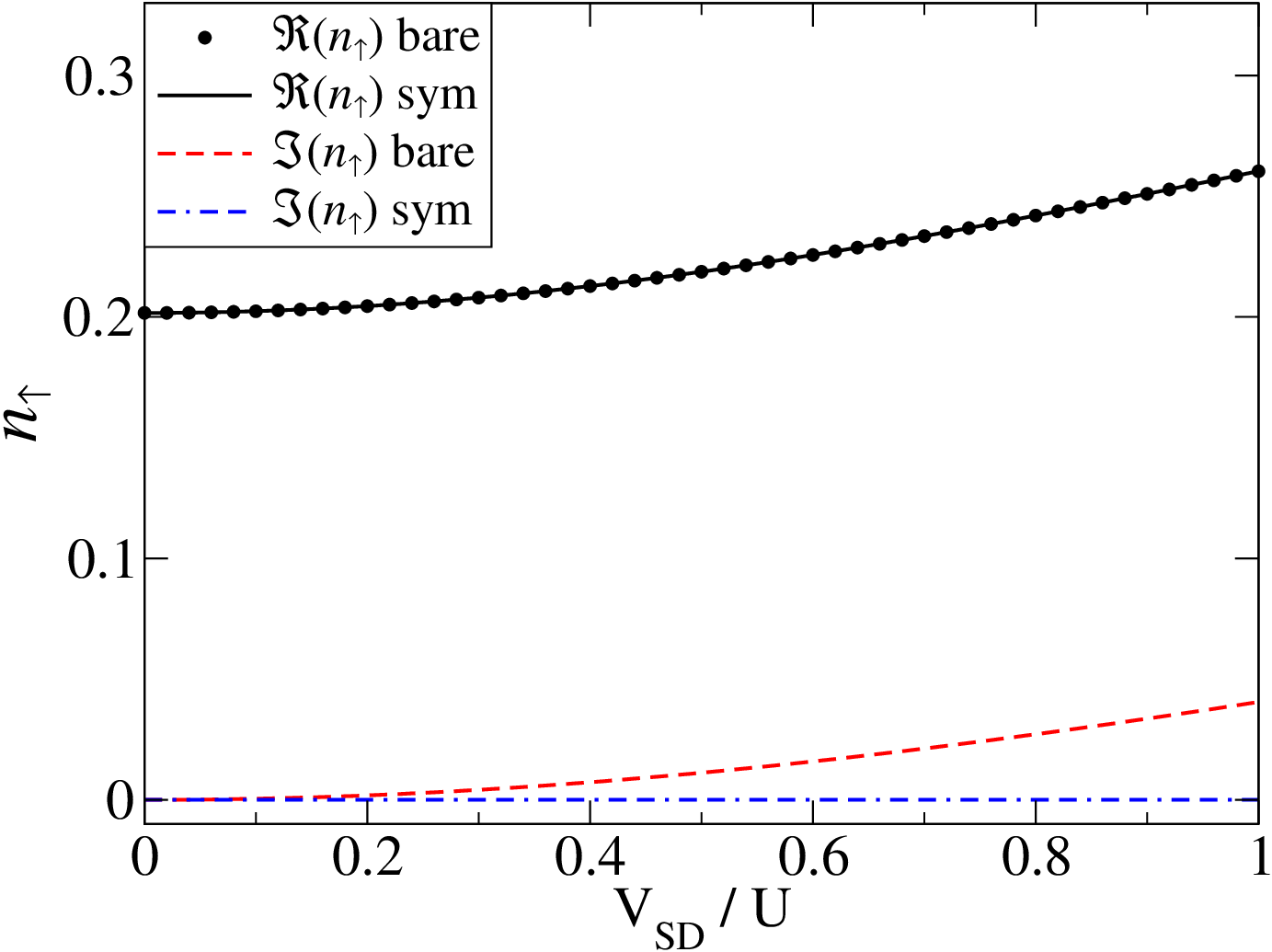}\caption{\label{fig:Occupation-of-the}Upper panel: Density of states in the
Kondo regime for nonequilibrium situation of the spin up electron
after symmetrization for different temperatures. Parameters used are
similar to those used in Refs.~\onlinecite{Meir1993,Galperin2007a}
(in units of $\Gamma=\Gamma_{L}+\Gamma_{R}$): $\mu_{L}=3/10$, $\mu_{R}=0$,
$\varepsilon_{\downarrow,\uparrow}=-2$, and $U=10$. The bands are
modeled as a Lorentzian with a half bandwidth $100$. Lower panel:
Occupation of the spin up electron before (``bare'') and after (``sym'')
symmetrization. As can be clearly seen the real part of the observable
$\left\langle n_{\uparrow}\right\rangle $ is not affected by the
symmetrization, and only the non physical imaginary part disappears.
Parameters used (in units of $U$): $\Gamma_{L,\uparrow}=\Gamma_{R,\uparrow}=0.3$,
$\Gamma_{L,\downarrow}=\Gamma_{R,\downarrow}=0.05$, $\varepsilon_{\uparrow}=0.2$,
$\varepsilon_{\downarrow}=-0.2$, $\beta=4$ and $U=1$.}
\end{figure}

First, we address the effects of symmetry breakage in the Anderson
model. The closure used is sufficient to describe the appearance of
the Kondo resonances at low temperatures, as seen in the upper panel
of figure \ref{fig:Occupation-of-the}, where we plot the density
of states as a function of energy for several temperatures, all calculated
with symmetry restoration. The development of Kondo peaks in the density
of states as the temperature decreases is clearly evident, signifying
a regime of strong correlations which is qualitatively captured by
the simple EOM approach when symmetry is restored. 

In the lower panel of figure \ref{fig:Occupation-of-the} we show
one of the main flaws of the EOM approach for the Anderson impurity
model, where we plot the value of $\left\langle n_{\uparrow}\right\rangle $
as a function of the source drain bias voltage with and without symmetry
restoration. The most notable effect is the appearance of an imaginary
portion to $\left\langle n_{\uparrow}\right\rangle $ as the source
drain bias voltage is increased. To obtain the results, without symmetry
restoration, only the real part of $\left\langle n_{\sigma}\right\rangle $
was used to converge the self-consistent equations for the GFs. By
applying the symmetrization scheme proposed in Sec. \ref{sub:Symmetry-restoration}
to the lesser GF calculated in Sec. \ref{sub:The-Anderson-model},
we restore the relation $G_{\alpha\beta}^{<,>}\left(\omega\right)=-\left(G_{\beta\alpha}^{<,>}\left(\omega\right)\right)^{*}$.
This is sufficient to obtain a real value for $\left\langle n_{\uparrow}\right\rangle $,
as clearly shown in the lower panel of figure \ref{fig:Occupation-of-the}.
All other symmetry relation are not violated here and thus, our symmetrization
procedure does not affect them at all. Interestingly, taking only
the real part of $\left\langle n_{\sigma}\right\rangle $ provides
identical results when compared to the results obtained after the
full symmetrization procedure. However, this is only true for the
simple case of the single site impurity model and does not hold for
more complex systems. 

\subsection{The double Anderson model\label{sub:Symmetrization-of-the-DA}}

We now turn to discuss the impact of symmetry breaking for the double
Anderson model. This system is more involved compared to the single
site Anderson model and thus, the level of closure used is somewhat
simpler, as explained in Sec. \ref{sub:The-double-Anderson}. While
for the case of a single site Anderson model only the relation $G_{\alpha\beta}^{<,>}\left(\omega\right)=-\left(G_{\beta\alpha}^{<,>}\left(\omega\right)\right)^{*}$
breaks down, in the double Anderson model we find that all $3$ symmetries
described by equation \ref{eq:relations} are violated. This can be
traced to the more complex form of the Hamiltonian for the double
Anderson model, where each site is only coupled to one of the leads
and transport in enabled by the direct hopping term between the two
sites. 

\begin{figure}
\centering{}\includegraphics[width=8cm]{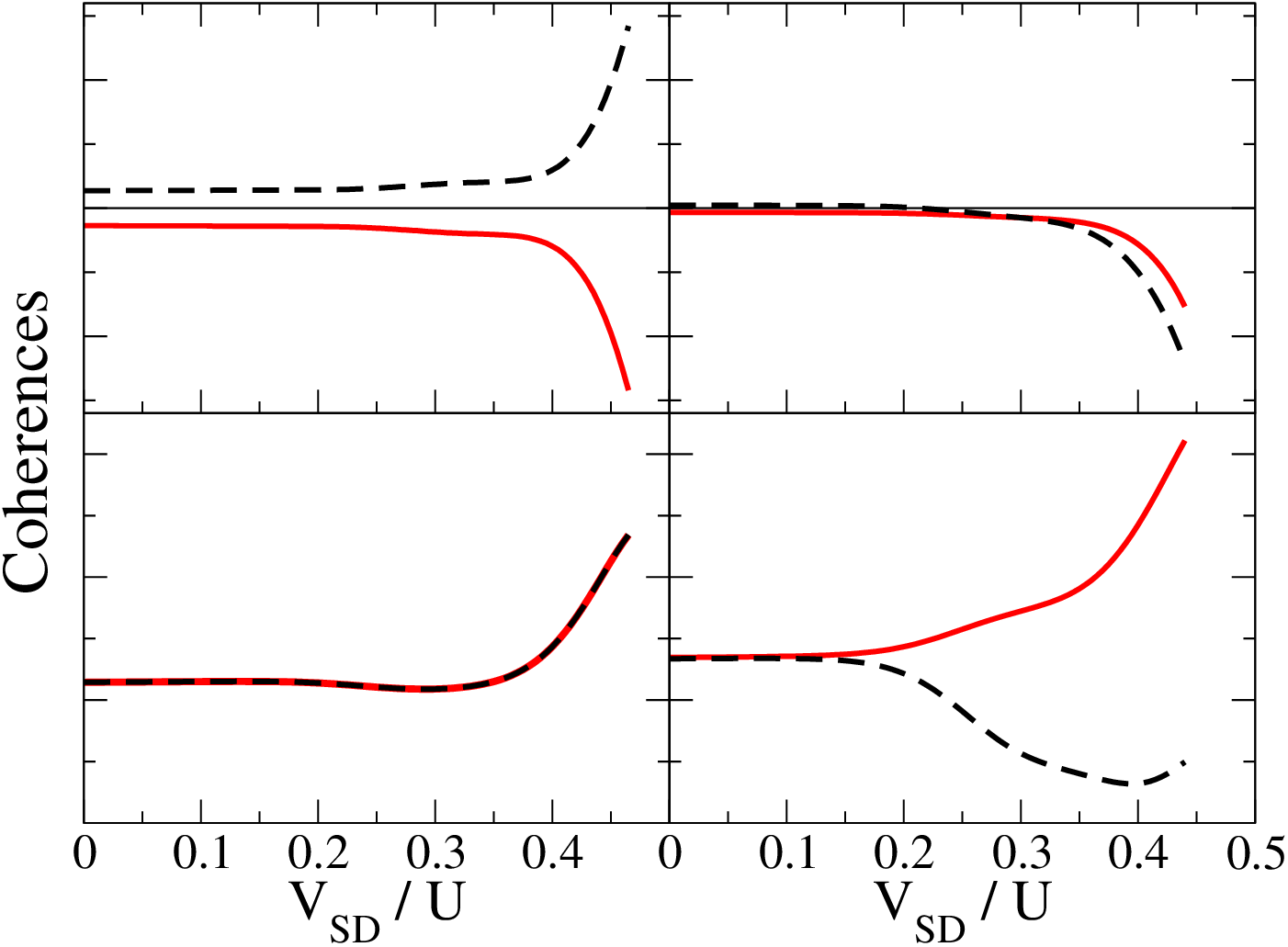}\caption{\label{fig:coherences}The imaginary (upper panels) and real (lower
panels) parts of $\rho_{\alpha\beta}^{\sigma\sigma}=\left\langle d_{\beta\sigma}^{\dagger}d_{\alpha\sigma}\right\rangle $
(dashed line) and $\rho_{\beta\alpha}^{\sigma\sigma}=\left\langle d_{\alpha\sigma}^{\dagger}d_{\beta\sigma}\right\rangle $
(solid line) calculated before (right panels) and after (left panels)
symmetry was restored. The solid thin line in the upper panels marks
the zero axis. As expected, after symmetry restoration (left panels),
$\Im\left(\rho_{\alpha\beta}^{\sigma\sigma}\right)=-\Im\left(\rho_{\beta\alpha}^{\sigma\sigma}\right)$
and $\Re\left(\rho_{\alpha\beta}^{\sigma\sigma}\right)=\Re\left(\rho_{\beta\alpha}^{\sigma\sigma}\right)$,
while before symmetry restoration (right panels) these equalities
are violated. Parameters used for the simulations in units of $U=U_{\alpha}=U_{\beta}$
are: $\Gamma_{\alpha\uparrow}^{L}=\Gamma_{\alpha\downarrow}^{L}=\Gamma_{\beta\uparrow}^{R}=\Gamma_{\beta\downarrow}^{R}=0.0025$,
$\Gamma_{\alpha\uparrow}^{R}=\Gamma_{\alpha\downarrow}^{R}=\Gamma_{\beta\uparrow}^{L}=\Gamma_{\beta\downarrow}^{L}=0$,
$h_{\alpha\beta}^{\sigma}=\bar{h_{\alpha\beta}^{\sigma}}=0.25$, $V_{\alpha\beta}^{\sigma\tau}=0.1$,
$\varepsilon_{\alpha\uparrow}=\varepsilon_{\alpha\downarrow}=0.1$,
$\varepsilon_{\beta\uparrow}=\varepsilon_{\beta\downarrow}=-0.175$
and $\beta=80$.}
\end{figure}

Similar to the case of the Anderson model, as a result of symmetry
breaking the occupation of the levels $\left\langle n_{\alpha\sigma}\right\rangle $
is a complex number. In addition, the coherences, $\rho_{\alpha\beta}^{\sigma\sigma}=-\frac{i\hbar}{2\pi}\int_{-\infty}^{\infty}\left(G_{\alpha\beta}^{\sigma\sigma}\left(\omega\right)\right)^{<}\mbox{d}\omega$,
should also fulfill certain symmetry relations, such as $\rho_{\alpha\beta}^{\sigma\sigma}=\left(\rho_{\beta\alpha}^{\sigma\sigma}\right)^{*}$.
In figure \ref{fig:coherences} we plot the real and imaginary parts
of $\rho_{\alpha\beta}^{\sigma\sigma}$ and $\rho_{\beta\alpha}^{\sigma\sigma}$
for the case where the symmetry procedure has been applied (left panels)
and for the bare case (right panels). The upper panels show the imaginary
part of $\rho_{\alpha\beta}^{\sigma\sigma}$ and $\rho_{\beta\alpha}^{\sigma\sigma}$,
which should show a mirror reflection about the zero axis (shown as
thin solid line). This is, indeed, the case when symmetry is restored,
however, it is destroyed when symmetry breaks down, in particular
as the source drain bias increases. A more dramatic effect is shown
for the real part of $\rho_{\alpha\beta}^{\sigma\sigma}$ and $\mathbf{\rho_{\beta\alpha}^{\sigma\sigma}}$
(lower panels). The two curves representing $\Re\left(\rho_{\alpha\beta}^{\sigma\sigma}\right)$
and $\Re\left(\rho_{\beta\alpha}^{\sigma\sigma}\right)$ should be
identical (left panel when symmetry is restored) but are quite distinct
when symmetry is not obeyed (right panel).

\begin{figure}
\centering{}\includegraphics[width=8cm]{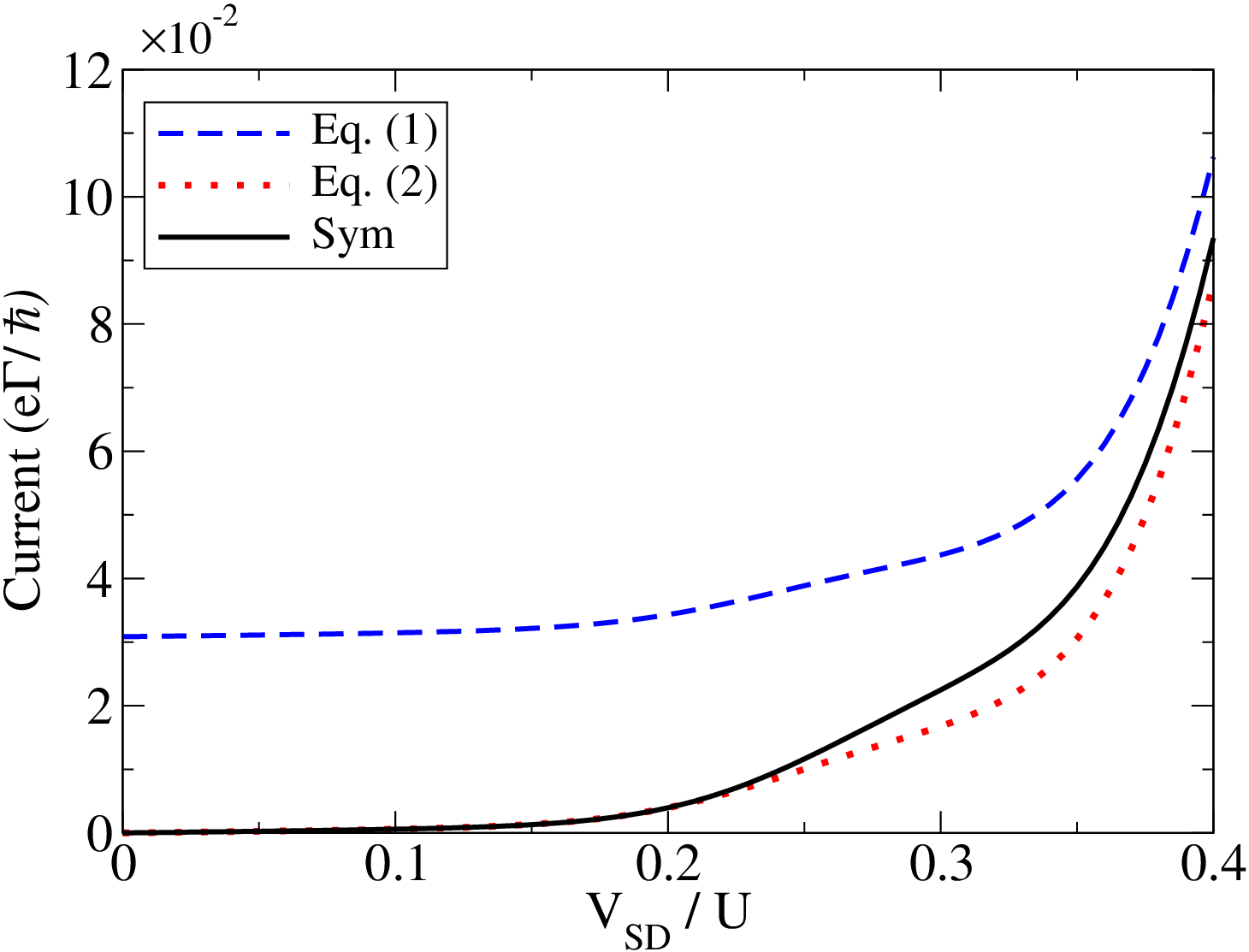}\caption{\label{fig:current}I-V curves calculated using equations (\ref{eq:I1})
and (\ref{eq:I2}) before (dashed and dotted lines) and after (solid
line) applying the symmetry procedure suggested in Sec. \ref{sub:Symmetry-restoration}.
As can be clearly seen, before symmetrization, calculating the current
via the two different but equivalent formulas provide different results,
one of which is not physical (dashed line) as the current is finite
for $V_{SD}=0.$ The latter result suggests that the ``unsymmetrized''
GFs obtained through the EOM disobey the fluctuation dissipation relation.
Parameters used for the simulations in units of $U=U_{\alpha}=U_{\beta}$
are: $\Gamma_{\alpha\uparrow}^{L}=\Gamma_{\alpha\downarrow}^{L}=\Gamma_{\beta\uparrow}^{R}=\Gamma_{\beta\downarrow}^{R}=0.0025$,
$\Gamma_{\alpha\uparrow}^{R}=\Gamma_{\alpha\downarrow}^{R}=\Gamma_{\beta\uparrow}^{L}=\Gamma_{\beta\downarrow}^{L}=0$,
$h_{\alpha\beta}^{\sigma}=\bar{h_{\alpha\beta}^{\sigma}}=0.25$, $V_{\alpha\beta}^{\sigma\tau}=0.1$,
$\varepsilon_{\alpha\uparrow}=\varepsilon_{\alpha\downarrow}=0.1$,
$\varepsilon_{\beta\uparrow}=\varepsilon_{\beta\downarrow}=-0.175$
and $\beta=80$.}
\end{figure}

In figure \ref{fig:current} we plot the current as a function of
the source drain bias voltage for the double Anderson model. The current
can be obtained from equation (\ref{eq:I1}) (dashed line) or from
equation (\ref{eq:I2}) (dotted curve). In the limit of infinite hierarchy
in the EOM approach the two formulas should coincide. However, when
approximations are introduced or when the hierarchy is truncated,
the calculation of the current based on the two different formulas
will coincide only if the symmetry relation $\left(G_{\alpha\beta}^{\sigma\tau}\left(\omega\right)\right)^{r}-\left(G_{\alpha\beta}^{\sigma\tau}\left(\omega\right)\right)^{a}=\left(G_{\alpha\beta}^{\sigma\tau}\left(\omega\right)\right)^{>}-\left(G_{\alpha\beta}^{\sigma\tau}\left(\omega\right)\right)^{<}$
is preserved. Indeed, in the case of a single site Anderson model,
even if symmetry is not restored, this relation holds and the two
calculations yield identical values for the current. However, in the
present case, all $3$ symmetry relations are broken and thus, equations
(\ref{eq:I1}) and (\ref{eq:I2}) give different results for the current,
as clearly evident in figure \ref{fig:current}. More significantly
is the fact that equation (\ref{eq:I1}) produces a finite value for
the current even when the bias is zero, indicating the break down
of the fluctuation dissipation relation. When symmetry is restored
(solid curve) the two calculations are identical, as they should be,
and the violation of the fluctuation dissipation relation is also
resolved.

\begin{figure}
\centering{}\includegraphics[width=8cm]{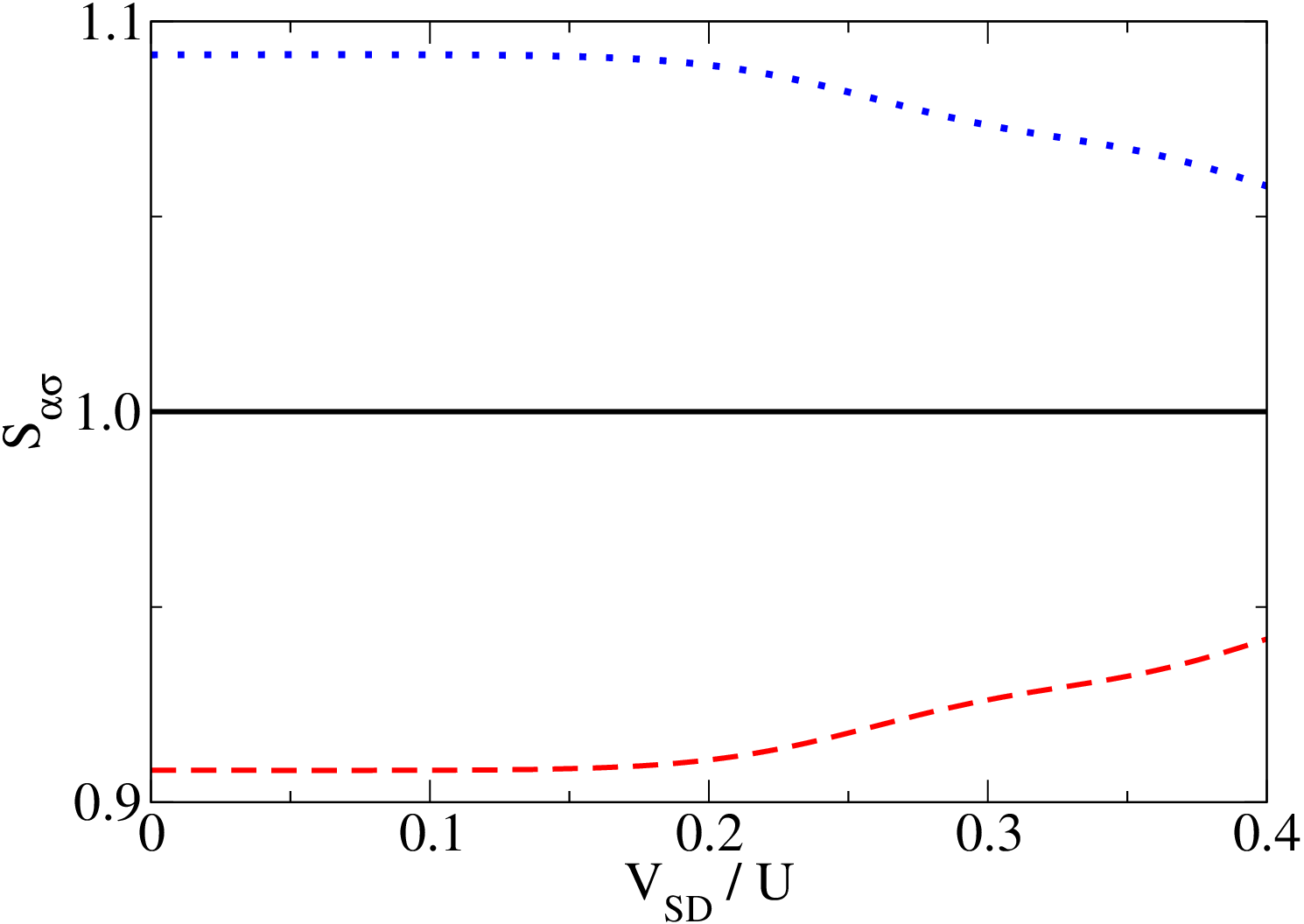}\caption{\label{fig:sum}$S_{\alpha\sigma}$ (dashed line) and $S_{\beta\sigma}$
(dotted line) calculated from the ``symmetrized'' lesser and greater
GFs as a function of the source drain bias voltage. The exact result
should have been 1 (as marked by the solid line). Parameters used
for the simulations in units of $U=U_{\alpha}=U_{\beta}$ are: $\Gamma_{\alpha\uparrow}^{L}=\Gamma_{\alpha\downarrow}^{L}=\Gamma_{\beta\uparrow}^{R}=\Gamma_{\beta\downarrow}^{R}=0.0025$,
$\Gamma_{\alpha\uparrow}^{R}=\Gamma_{\alpha\downarrow}^{R}=\Gamma_{\beta\uparrow}^{L}=\Gamma_{\beta\downarrow}^{L}=0$,
$h_{\alpha\beta}^{\sigma}=\bar{h_{\alpha\beta}^{\sigma}}=0.25$, $V_{\alpha\beta}^{\sigma\tau}=0.1$,
$\varepsilon_{\alpha\uparrow}=\varepsilon_{\alpha\downarrow}=0.1$,
$\varepsilon_{\beta\uparrow}=\varepsilon_{\beta\downarrow}=-0.175$
and $\beta=80$.}
\end{figure}

The symmetrization scheme proposed here is not a ``magic cure''
and, in fact, does not resolve all issues of mater. It is well known
that the lesser and greater GFs should obey a simple sum rule where
the integral over the difference of their diagonal elements should
always sum to $1$: 
\begin{equation}
S_{\alpha\sigma}=\frac{-i\hbar}{2\pi}\int\mbox{d}\varepsilon\left(\left(G_{\alpha\alpha}^{\sigma\sigma}\left(\epsilon\right)\right)^{<}-\left(G_{\alpha\alpha}^{\sigma\sigma}\left(\epsilon\right)\right)^{>}\right)=1.\label{eq:sum}
\end{equation}
In figure \ref{fig:sum} we plot the sum rule as given by equation
(\ref{eq:sum}) for the double Anderson model where symmetry has been
restored. A similar plot for the single site Anderson model yields
a value of $1$ regardless of whether symmetry has been restored or
not within the closure discussed above. However, in the case of the
more evolved double Anderson model, even when symmetry is restored
and the GFs obey all $3$ relations described in equation (\ref{eq:relations}),
the sum rule is violated. Nonetheless, the sum $\underset{\alpha\sigma}{\sum}S_{\alpha\sigma}=N_{e}$,
where $N_{e}$ is the total number of electrons in the system at maximal
occupancy, is indeed preserved when symmetrization is restored. 

\section{Summary\label{sec:Conclusions-and-Summary}}

In this paper we have addressed the problem of symmetry breaking and
restoring in the EOM approach to NEGF formalism. This formalism is
based on deriving a hierarchy of equations of motion for the system's
Green functions and truncating this hierarchy at a desired (or tractable)
order. Despite the uncontrolled approximation introduced by an arbitrary
truncation, the closed set of equations is often used to describe
the complex dynamics of correlated systems, including the Coulomb
blockade and Kondo effect. 

One shortcoming of the EOM approach, which has been the focus of the
present study, is the fact that, a priori, for most situations it
is impossible to determine whether the solution of the closed set
of equations satisfies symmetry relation between the retarded, advanced,
lesser and greater Green functions imposed by definition. For example,
we have shown that for the Anderson model the relation $G_{\alpha\beta}^{<,>}\left(\omega\right)=-\left(G_{\beta\alpha}^{<,>}\left(\omega\right)\right)^{*}$
breaks down for a closure that is often used to describe the dynamics
near the Kondo regime. We have also demonstrated that for the double
Anderson model all $3$ symmetry relations given by equation (\ref{eq:relations})
break down for a lower level of closure. This faulty of the EOM approach
leads to unphysical behavior such as complex level occupations and
finite current at zero source drain bias (depending on how the current
is evaluated). 

We have also proposed a procedure to circumvent this deficiency by
imposing symmetrization to the Green functions in such a way that
all $3$ symmetry relations are restored. The strength of the proposed
approach is that it does not alter the GFs if symmetry is not broken.
While this procedure eliminates some problems of physical importance
and leads to real level occupations and vanishing current at zero
source drain bias (irrespective of how the current is evaluated),
certain sum rules are still violated, indicating other problems with
the EOM approach. Nonetheless, the symmetrized version of the EOM
technique still describes the appearance of the Kondo peak and, as
will be shown in future publication provides a quantitative description
of the resonant transport for the double Anderson model even in the
strong inter-dot coupling limit.

\section{Acknowledgments\label{sec:Acknowledgments} }

We would like to thank Guy Cohen, Yigal Meir, Andrew Millis, Abe Nitzan,
David Reichman, and Eli Wilner for fruitful discussions. This work
was supported by the US-Israel Binational Science Foundation and by
the FP7 Marie Curie IOF project HJSC. TJL is grateful to the The Center
for Nanoscience and Nanotechnology at Tel Aviv University of a doctoral
fellowship. 

\bibliographystyle{achemso} \addcontentsline{toc}{section}{\refname}\bibliographystyle{apsrev}
\bibliography{Sym_Break}

\newpage{}

\begin{widetext}

\begin{center}
\textbf{\Large Supporting Information}
\par\end{center}{\Large \par}

\section{Full derivation of the broken symmetry in the Anderson model}

Here we present in greater detail the breakage of the relation $G_{\sigma\sigma}^{<}\left(\omega\right)=-\left(G_{\sigma\sigma}^{<}\left(\omega\right)\right)^{*}$
described in subsection \textbf{The Anderson model} in the manuscript.
We demonstrate that $G_{\sigma\sigma}^{<}\left(\omega\right)$ is
not an imaginary function. We start by defining the following contour
ordered GF:
\begin{equation}
G_{\sigma\sigma}\left(t,t'\right)=-\frac{i}{\hbar}\left\langle T_{C}d_{\sigma}\left(t\right)d_{\sigma}^{\dagger}\left(t'\right)\right\rangle ,
\end{equation}
\begin{equation}
G_{2}\left(t,t'\right)=-\frac{i}{\hbar}\left\langle T_{C}n_{\bar{\sigma}}\left(t\right)d_{\sigma}\left(t\right)d_{\sigma}^{\dagger}\left(t'\right)\right\rangle ,
\end{equation}
where $\bar{\sigma}$ is the opposite spin of $\sigma$. The resulting
EOMs (in Fourier space) under the approximation scheme discuss in
the manuscript are: 
\begin{equation}
\left(\hbar\omega-\varepsilon_{\sigma}-\Sigma_{0}\left(\omega\right)\right)G_{\sigma\sigma}\left(\omega\right)=1+UG_{2}\left(\omega\right),
\end{equation}
\begin{eqnarray}
G_{2}\left(\omega\right) & = & \left(\hbar\omega-\varepsilon_{\sigma}-U-\Sigma_{0}\left(\omega\right)-\Sigma_{3}\left(\omega\right)\right)^{-1}\left(\left\langle n_{\bar{\sigma}}\right\rangle -\Sigma_{1}\left(\omega\right)G_{\sigma\sigma}\left(\omega\right)\right),
\end{eqnarray}
We define the following GFs and self-energies: 
\begin{equation}
g\left(\omega\right)=\frac{1}{\hbar\omega-\varepsilon_{\sigma}-\Sigma_{0}\left(\omega\right)},
\end{equation}
\begin{equation}
g_{2}\left(\omega\right)=\frac{1}{\hbar\omega-\varepsilon_{\sigma}-U-\Sigma_{4}\left(\omega\right)},
\end{equation}
\begin{equation}
\Sigma_{0}\left(\omega\right)=\sum_{i,k\in\left\{ L,R\right\} }\frac{\left|t_{k\sigma}\right|^{2}}{\hbar\omega-\varepsilon_{k,i,\sigma}},\label{eq:selfenergy}
\end{equation}
\begin{eqnarray}
\Sigma_{j}\left(\omega\right) & = & \sum_{i,k\in\left\{ L,R\right\} }A_{i,k}^{\left(j\right)}\left|t_{k\sigma}\right|^{2}\left(\frac{1}{\hbar\omega+\varepsilon_{k,i,\bar{\sigma}}-\varepsilon_{\sigma}-\varepsilon_{\bar{\sigma}}-U}+\frac{1}{\hbar\omega-\varepsilon_{k,i,\bar{\sigma}}-\varepsilon_{\sigma}+\varepsilon_{\bar{\sigma}}}\right),\, j=1,3\nonumber \\
\end{eqnarray}
\begin{equation}
\Sigma_{4}\left(\omega\right)=\Sigma_{0}\left(\omega\right)+\Sigma_{3}\left(\omega\right).
\end{equation}
with $A_{k}^{\left(1\right)}=f_{k}\left(\varepsilon_{i,k,\sigma}-\mu_{k}\right)$,
$A_{k}^{\left(3\right)}=1$, and $f_{k}\left(\varepsilon_{i,k,\sigma}-\mu_{k}\right)$
is the Fermi Dirac distribution. 

Rewriting the equations of the GFs in terms of the above definitions
gives:
\begin{equation}
G_{2}\left(\omega\right)=g_{2}\left(\omega\right)\left\langle n_{\bar{\sigma}}\right\rangle -g_{2}\left(\omega\right)\Sigma_{1}\left(\omega\right)G_{\sigma\sigma}\left(\omega\right),\label{eq:G2}
\end{equation}
\begin{equation}
G_{\sigma\sigma}\left(\omega\right)=g\left(\omega\right)+g\left(\omega\right)UG_{2}\left(\omega\right),\label{eq:G}
\end{equation}
We then merge equations (\ref{eq:G}) and (\ref{eq:G2}) to get: 
\begin{equation}
G_{2}\left(\omega\right)=g_{2}\left(\omega\right)\left\langle n_{\bar{\sigma}}\right\rangle -g_{2}\left(\omega\right)\Sigma_{1}\left(\omega\right)g\left(\omega\right)-g_{2}\left(\omega\right)\Sigma_{1}\left(\omega\right)g\left(\omega\right)UG_{2}\left(\omega\right).\label{eq:G22}
\end{equation}
The advanced GF can be extracted from the contour ordered one (equation
(\ref{eq:G22})) by setting~\citealp{Haug1996} $\omega\rightarrow\omega-i0^{+}$

\begin{eqnarray}
G_{2}^{a}\left(\omega\right) & = & g_{2}^{a}\left(\omega\right)\left\langle n_{\bar{\sigma}}\right\rangle -g_{2}^{a}\left(\omega\right)\Sigma_{1}^{a}\left(\omega\right)g^{a}\left(\omega\right)-g_{2}^{a}\left(\omega\right)\Sigma_{1}^{a}\left(\omega\right)g^{a}\left(\omega\right)UG_{2}^{a}\left(\omega\right),\label{eq:13}
\end{eqnarray}
For brevity, we omit $\left(\omega\right)$ and rewrite equation (\ref{eq:13})
as: 
\begin{eqnarray}
G_{2}^{a} & = & \left(1+g_{2}^{a}\Sigma_{1}^{a}g^{a}U\right)^{-1}g_{2}^{a}\left\langle n_{\bar{\sigma}}\right\rangle -\left(1+g_{2}^{a}\Sigma_{1}^{a}g^{a}U\right)^{-1}g_{2}^{a}\Sigma_{1}^{a}g^{a}.
\end{eqnarray}
Define:
\begin{equation}
P^{r,a}=\left(1+g_{2}^{r,a}\Sigma_{1}^{r,a}g^{r,a}U\right)^{-1},
\end{equation}
\begin{eqnarray}
G_{2}^{a} & = & P^{a}g_{2}^{a}\left\langle n_{\bar{\sigma}}\right\rangle -P^{a}g_{2}^{a}\Sigma_{1}^{a}g^{a},
\end{eqnarray}
Using Langreth theorem, the lesser projection of $G_{2}$ can be evaluated:
\begin{equation}
G_{2}^{<}=g_{2}^{<}\left\langle n_{\bar{\sigma}}\right\rangle -\left(g_{2}\Sigma_{1}g\right)^{<}-\left(g_{2}\Sigma_{1}g\right)^{r}UG_{2}^{<}-\left(g_{2}\Sigma_{1}g\right)^{<}UG_{2}^{a},\label{eq:G2l}
\end{equation}
with 
\begin{equation}
g^{<}=g^{r}\Sigma_{0}^{<}g^{a},\label{eq:gl}
\end{equation}
\begin{equation}
g_{2}^{<}=g_{2}^{r}\Sigma_{4}^{<}g_{2}^{a},\label{eq:g2l}
\end{equation}
and the lesser self energies are defined as in Ref.~\onlinecite{Galperin2007a}:
\begin{equation}
\Sigma_{x}^{<}=\Sigma_{xL}^{<}+\Sigma_{xR}^{<}=i\left(f_{L}\Gamma_{xL}+f_{R}\Gamma_{xR}\right),\label{eq:lessSelfEnr1}
\end{equation}
where
\begin{equation}
\Gamma_{xk}=-2Im\left(\Sigma_{xk}^{r}\right),\label{eq:lessSelfEnr2}
\end{equation}
and $\Sigma_{xk}^{r,a}$ stands for the retarded (``r'') or advanced
(``a'') self-energies. Substituting equations (\ref{eq:gl}) and
(\ref{eq:g2l}) into equation (\ref{eq:G2l}), the lesser projection
of equation (\ref{eq:G22}) is given by: 
\begin{eqnarray}
G_{2}^{<} & = & P^{r}g_{2}^{<}\left\langle n_{\bar{\sigma}}\right\rangle -P^{r}g_{2}^{r}\Sigma_{1}^{r}g^{<}-P^{r}g_{2}^{r}\Sigma_{1}^{<}g^{a}-P^{r}g_{2}^{<}\Sigma_{1}^{a}g^{a}\nonumber \\
 &  & -P^{r}g_{2}^{r}\Sigma_{1}^{r}g^{<}UP^{a}g_{2}^{a}\left\langle n_{\bar{\sigma}}\right\rangle -P^{r}g_{2}^{r}\Sigma_{1}^{<}g^{a}UP^{a}g_{2}^{a}\left\langle n_{\bar{\sigma}}\right\rangle -P^{r}g_{2}^{<}\Sigma_{1}^{a}g^{a}UP^{a}g_{2}^{a}\left\langle n_{\bar{\sigma}}\right\rangle \nonumber \\
 &  & +P^{r}g_{2}^{r}\Sigma_{1}^{r}g^{<}UP^{a}g_{2}^{a}\Sigma_{1}^{a}g^{a}+P^{r}g_{2}^{r}\Sigma_{1}^{<}g^{a}UP^{a}g_{2}^{a}\Sigma_{1}^{a}g^{a}+P^{r}g_{2}^{<}\Sigma_{1}^{a}g^{a}UP^{a}g_{2}^{a}\Sigma_{1}^{a}g^{a},
\end{eqnarray}
The lesser projection of equation (\ref{eq:G}) can now be written
as 
\begin{equation}
G_{\sigma\sigma}^{<}=g^{<}+g^{r}UG_{2}^{<}+g^{<}UG_{2}^{a}.
\end{equation}
Using our results for $G_{2}^{a}$ and $G_{2}^{<}$ we find
\begin{eqnarray}
G_{\sigma\sigma}^{<} & = & g^{<}+g^{r}UP^{r}g_{2}^{<}\left\langle n_{\bar{\sigma}}\right\rangle +g^{<}UP^{a}g_{2}^{a}\left(\left\langle n_{\bar{\sigma}}\right\rangle -\Sigma_{1}^{a}g^{a}\right)-g^{r}UP^{r}g_{2}^{<}\Sigma_{1}^{a}g^{a}\nonumber \\
 &  & -g^{r}UP^{r}g_{2}^{r}\left(\Sigma_{1}^{r}g^{<}+\Sigma_{1}^{<}g^{a}\right)-g^{r}UP^{r}g_{2}^{r}\Sigma_{1}^{r}g^{<}UP^{a}g_{2}^{a}\left(\left\langle n_{\bar{\sigma}}\right\rangle +\Sigma_{1}^{a}g^{a}\right)\\
 &  & -g^{r}UP^{r}g_{2}^{r}\Sigma_{1}^{<}g^{a}UP^{a}g_{2}^{a}\left(\left\langle n_{\bar{\sigma}}\right\rangle +\Sigma_{1}^{a}g^{a}\right)-g^{r}UP^{r}g_{2}^{<}\Sigma_{1}^{a}g^{a}UP^{a}g_{2}^{a}\left(\left\langle n_{\bar{\sigma}}\right\rangle +\Sigma_{1}^{a}g^{a}\right)\nonumber 
\end{eqnarray}
Applying the principle of reductio ad absurdum, we assume $G_{\sigma\sigma}^{<}$
is imaginary. Since it must hold for any real value of $\left\langle n_{\bar{\sigma}}\right\rangle $
between $0$ and $1$, we argue that the term 
\begin{eqnarray}
A_{1} & = & g^{r}UP^{r}g_{2}^{<}\left\langle n_{\bar{\sigma}}\right\rangle +g^{<}UP^{a}g_{2}^{a}\left\langle n_{\bar{\sigma}}\right\rangle -g^{r}UP^{r}g_{2}^{r}\Sigma_{1}^{r}g^{<}UP^{a}g_{2}^{a}\left\langle n_{\bar{\sigma}}\right\rangle \nonumber \\
 &  & -g^{r}UP^{r}g_{2}^{r}\Sigma_{1}^{<}g^{a}UP^{a}g_{2}^{a}\left\langle n_{\bar{\sigma}}\right\rangle -g^{r}UP^{r}g_{2}^{<}\Sigma_{1}^{a}g^{a}UP^{a}g_{2}^{a}\left\langle n_{\bar{\sigma}}\right\rangle ,
\end{eqnarray}
is imaginary by itself. Moreover, Since $A_{1}$ must be imaginary
for any value of $U$, the term 
\begin{eqnarray}
A_{2} & = & g^{r}UP^{r}g_{2}^{<}\left\langle n_{\bar{\sigma}}\right\rangle +g^{<}UP^{a}g_{2}^{a}\left\langle n_{\bar{\sigma}}\right\rangle ,
\end{eqnarray}
should be imaginary as well. Using the fact that $U$ and $\left\langle n_{\bar{\sigma}}\right\rangle $
are real quantities and by definition $g_{2}^{<}$ and $g^{<}$ are
imaginary, for $A_{2}$ to be imaginary, one requires that its real
part vanishes, i.e.,:
\begin{equation}
Im\left(g^{r}P^{r}\right)g_{2}^{<}+Im\left(P^{a}g_{2}^{a}\right)g^{<}=0.\label{eq:demand1-1}
\end{equation}
In other words the equality
\begin{equation}
Im\left(g^{r}P^{r}\right)g_{2}^{r}\Sigma_{4}^{<}g_{2}^{a}=-Im\left(P^{a}g_{2}^{a}\right)g^{r}\Sigma_{0}^{<}g^{a},
\end{equation}
must hold for the assumption that $G_{\sigma\sigma}^{<}$ is imaginary
to be satisfied. Using the definitions for $g$ and $g_{2}$ the last
equality can be rewritten as: 
\begin{equation}
Im\left(g^{r}P^{r}\right)\frac{-if\left(\omega\right)Im\left(\Sigma_{4}^{r}\right)}{\left(\hbar\omega-\varepsilon_{4}-U\right)^{2}+\left(Im\left(\Sigma_{4}^{r}\right)\right)^{2}}=Im\left(P^{a}g_{2}^{a}\right)\frac{if\left(\omega\right)Im\left(\Sigma_{0}^{r}\right)}{\left(\hbar\omega-\varepsilon_{0}\right)^{2}+\left(Im\left(\Sigma_{0}^{r}\right)\right)^{2}},\label{eq:equ}
\end{equation}
where $\varepsilon_{0}=\varepsilon_{\sigma}+Re\left(\Sigma_{0}^{r}\right)$
and $\varepsilon_{4}=\varepsilon_{\sigma}+Re\left(\Sigma_{4}^{r}\right)$.
Starting with the L.H.S. of equation (\ref{eq:equ}), we look at $g^{r}P^{r}$
\begin{eqnarray}
g^{r}P^{r} & = & \frac{g^{r}}{1+g_{2}^{r}\Sigma_{1}^{r}g^{r}U}=\frac{1}{\hbar\omega-\varepsilon_{\sigma}-\Sigma_{0}^{r}+\frac{\Sigma_{1}^{r}U}{\hbar\omega-\varepsilon_{\sigma}-U-\Sigma_{4}^{r}}}\nonumber \\
 & = & \frac{\hbar\omega-\varepsilon_{\sigma}-U-\Sigma_{4}^{r}}{\left(\hbar\omega-\varepsilon_{\sigma}-\Sigma_{0}^{r}\right)\left(\hbar\omega-\varepsilon_{\sigma}-U-\Sigma_{4}^{r}\right)+\Sigma_{1}^{r}U}\\
 & = & \frac{\hbar\omega-\varepsilon_{4}-U-i\left(Im\Sigma_{4}^{r}\right)}{\left(\hbar\omega-\varepsilon_{0}-i\left(Im\Sigma_{0}^{r}\right)\right)\left(\hbar\omega-\varepsilon_{4}-U-i\left(Im\Sigma_{4}^{r}\right)\right)+U\cdot Re\left(\Sigma_{1}^{r}\right)+U\cdot i\left(Im\Sigma_{1}^{r}\right)}.\nonumber 
\end{eqnarray}
Denote $a_{0}=\hbar\omega-\varepsilon_{0}$, $a_{1}=Re\left(\Sigma_{1}^{r}\right)$,
$a_{4}=\hbar\omega-\varepsilon_{4}-U$ and $b_{x}=Im\left(\Sigma_{x}^{r}\right)$
with $x=0,1,4$
\begin{eqnarray}
g^{r}P^{r} & = & \frac{a_{4}-ib_{4}}{\left(a_{0}-ib_{0}\right)\left(a_{4}-ib_{4}\right)+Ua_{1}+iUb_{1}}\nonumber \\
 & = & \frac{a_{4}-ib_{4}}{a_{0}a_{4}-b_{0}b_{4}+Ua_{1}-i\left(a_{0}b_{4}+a_{4}b_{0}-Ub_{1}\right)}.\label{eq:31}
\end{eqnarray}
Denote $D=a_{0}a_{4}-b_{0}b_{4}+Ua_{1}$ and $E=a_{0}b_{4}+a_{4}b_{0}-Ub_{1}$,
so we can rewrite equation (\ref{eq:31}) as 
\begin{eqnarray}
g^{r}P^{r} & = & \frac{a_{4}-ib_{4}}{D-iE}=\frac{\left(a_{4}-ib_{4}\right)\left(D+iE\right)}{D^{2}+E^{2}}=\frac{a_{4}D+b_{4}E-i\left(b_{4}D-a_{4}E\right)}{D^{2}+E^{2}}.
\end{eqnarray}
Finally
\begin{equation}
Im\left(g^{r}P^{r}\right)=-\frac{b_{4}D-a_{4}E}{D^{2}+E^{2}}
\end{equation}
The L.H.S. of equation (\ref{eq:equ}) is thus
\begin{equation}
Im\left(g^{r}P^{r}\right)\frac{-if\left(\omega\right)Im\left(\Sigma_{4}^{r}\right)}{\left(\hbar\omega-\varepsilon_{4}-U\right)^{2}+\left(Im\left(\Sigma_{4}^{r}\right)\right)^{2}}=\frac{if\left(\omega\right)b_{4}}{\left(a_{4}\right)^{2}+\left(b_{4}\right)^{2}}\frac{b_{4}D-a_{4}E}{D^{2}+E^{2}}.
\end{equation}
Now we turn to analyze the R.H.S. of equation (\ref{eq:equ}). We
start with evaluating $P^{a}g_{2}^{a}$ 
\begin{eqnarray}
P^{a}g_{2}^{a} & = & \frac{g_{2}^{a}}{1+g_{2}^{a}\Sigma_{1}^{a}g^{a}U}=\frac{1}{\hbar\omega-\varepsilon_{\sigma}-U-\Sigma_{4}^{a}+\frac{\Sigma_{1}^{a}U}{\hbar\omega-\varepsilon_{\sigma}-\Sigma_{0}^{a}}}\nonumber \\
 & = & \frac{\hbar\omega-\varepsilon_{\sigma}-\Sigma_{0}^{a}}{\left(\hbar\omega-\varepsilon_{\sigma}-U-\Sigma_{4}^{a}\right)\left(\hbar\omega-\varepsilon_{\sigma}-\Sigma_{0}^{a}\right)+\Sigma_{1}^{a}U}\nonumber \\
\nonumber \\
 & = & \frac{a_{0}+ib_{0}}{\left(a_{0}+ib_{0}\right)\left(a_{4}+ib_{4}\right)+Ua_{1}-iUb_{1}}\label{eq:last}\\
 & = & \frac{a_{0}+ib_{0}}{a_{0}a_{4}-b_{0}b_{4}+Ua_{1}+i\left(a_{0}b_{4}+a_{4}b_{0}-Ub_{1}\right)}\nonumber \\
 & = & \frac{a_{0}+ib_{0}}{D+iE}=\frac{\left(a_{0}+ib_{0}\right)\left(D-iE\right)}{D^{2}+E^{2}}=\frac{a_{0}D+b_{0}E+i\left(b_{0}D-a_{0}E\right)}{D^{2}+E^{2}}.\nonumber 
\end{eqnarray}
To go from the second line to the third line in equation (\ref{eq:last})
we used $Im\left(\Sigma_{x}^{r}\right)=-Im\left(\Sigma_{x}^{a}\right)$.
Finally:
\begin{equation}
Im\left(P^{a}g_{2}^{a}\right)=\frac{b_{0}D-a_{0}E}{D^{2}+E^{2}}.
\end{equation}
The R.H.S. of equation (\ref{eq:equ}) is thus,
\begin{equation}
Im\left(P^{a}g_{2}^{a}\right)\frac{if\left(\omega\right)Im\left(\Sigma_{0}^{r}\right)}{\left(\hbar\omega-\varepsilon_{0}\right)^{2}+\left(Im\left(\Sigma_{0}^{r}\right)\right)^{2}}=\frac{b_{0}D-a_{0}E}{D^{2}+E^{2}}\frac{if\left(\omega\right)b_{0}}{\left(a_{0}\right)^{2}+\left(b_{0}\right)^{2}}.
\end{equation}
The equality (equation(\ref{eq:equ})) now reads:
\begin{equation}
\frac{b_{4}}{\left(a_{4}\right)^{2}+\left(b_{4}\right)^{2}}\frac{b_{4}D-a_{4}E}{D^{2}+E^{2}}=\frac{b_{0}D-a_{0}E}{D^{2}+E^{2}}\frac{b_{0}}{\left(a_{0}\right)^{2}+\left(b_{0}\right)^{2}},
\end{equation}
or
\begin{equation}
\frac{b_{4}^{2}D-a_{4}b_{4}E}{\left(a_{4}\right)^{2}+\left(b_{4}\right)^{2}}=\frac{b_{0}^{2}D-a_{0}b_{0}E}{\left(a_{0}\right)^{2}+\left(b_{0}\right)^{2}}.
\end{equation}
Substituting $D=a_{0}a_{4}-b_{0}b_{4}+Ua_{1}$ and $E=a_{0}b_{4}+a_{4}b_{0}-Ub_{1}$
one can easily show that the equality \textbf{does not hold}. Thus,
$G_{\sigma\sigma}^{<}\left(\omega\right)$ is not an imaginary function
as it should be by definition.

In the paper we argued that a simpler closure (as used for example,
in Ref.~\onlinecite{Song2007}) will not violate the symmetries of
the GFs. In what follows we show that under the simpler closure, indeed,
$G_{\sigma\sigma}^{<}\left(\omega\right)=-\left(G_{\sigma\sigma}^{<}\left(\omega\right)\right)^{*}$.
Our starting point is the same. Define the contour ordered GFs:
\begin{equation}
G_{\sigma\sigma}\left(t,t'\right)=-\frac{i}{\hbar}\left\langle T_{C}d_{\sigma}\left(t\right)d_{\sigma}^{\dagger}\left(t'\right)\right\rangle ,
\end{equation}
\begin{equation}
G_{2}\left(t,t'\right)=-\frac{i}{\hbar}\left\langle T_{C}n_{\bar{\sigma}}\left(t\right)d_{\sigma}\left(t\right)d_{\sigma}^{\dagger}\left(t'\right)\right\rangle ,
\end{equation}
Following the approximations of Ref.~\onlinecite{Song2007}, the
resulting EOMs (in Fourier space) are: 
\begin{equation}
\left(\hbar\omega-\varepsilon_{\sigma}-\Sigma_{0}\left(\omega\right)\right)G_{\sigma\sigma}\left(\omega\right)=1+UG_{2}\left(\omega\right),\label{eq:1}
\end{equation}
\begin{eqnarray}
G_{2}\left(\omega\right) & = & \left(\hbar\omega-\varepsilon_{\sigma}-U-\Sigma_{0}\left(\omega\right)\right)^{-1}\left\langle n_{\bar{\sigma}}\right\rangle .\label{eq:2}
\end{eqnarray}
We define 
\begin{equation}
g\left(\omega\right)=\frac{1}{\hbar\omega-\varepsilon_{\sigma}-\Sigma_{0}\left(\omega\right)},
\end{equation}
\begin{equation}
g_{2}\left(\omega\right)=\frac{1}{\hbar\omega-\varepsilon_{\sigma}-U-\Sigma_{0}\left(\omega\right)},
\end{equation}
\begin{equation}
\Sigma_{0}\left(\omega\right)=\sum_{i,k\in\left\{ L,R\right\} }\frac{\left|t_{k\sigma}\right|^{2}}{\hbar\omega-\varepsilon_{k,i,\sigma}}.
\end{equation}
Rewriting equations (\ref{eq:1}) and (\ref{eq:2}) in terms of the
given definitions we get:
\begin{equation}
G_{2}\left(\omega\right)=g_{2}\left(\omega\right)\left\langle n_{\bar{\sigma}}\right\rangle ,\label{eq:G2-1}
\end{equation}
\begin{equation}
G_{\sigma\sigma}\left(\omega\right)=g\left(\omega\right)+g\left(\omega\right)UG_{2}\left(\omega\right),\label{eq:G-1}
\end{equation}
Substitute equation (\ref{eq:G2-1}) into equation (\ref{eq:G-1})
\begin{equation}
G_{\sigma\sigma}\left(\omega\right)=g\left(\omega\right)+g\left(\omega\right)U\left\langle n_{\bar{\sigma}}\right\rangle g_{2}\left(\omega\right).
\end{equation}
Applying the Langreth rules we find the lesser GF (omitting $\left(\omega\right)$
for brevity):
\begin{equation}
G_{\sigma\sigma}^{<}=g^{<}+g^{r}U\left\langle n_{\bar{\sigma}}\right\rangle g_{2}^{<}+g^{<}U\left\langle n_{\bar{\sigma}}\right\rangle g_{2}^{a}
\end{equation}
where $g^{<}=g^{r}\Sigma_{0}^{<}g^{a},$ $g_{2}^{<}=g_{2}^{r}\Sigma_{0}^{<}g_{2}^{a},$
and $\Sigma_{0}^{<}$ is defined in equations (\ref{eq:lessSelfEnr1})
and (\ref{eq:lessSelfEnr2}). By definition $g^{<}$ is imaginary,
hence, for $G_{\sigma\sigma}^{<}$ to be imaginary one requires that
\begin{eqnarray}
A_{1} & = & g^{r}U\left\langle n_{\bar{\sigma}}\right\rangle g_{2}^{r}\Sigma_{0}^{<}g_{2}^{a}+g^{r}\Sigma_{0}^{<}g^{a}U\left\langle n_{\bar{\sigma}}\right\rangle g_{2}^{a},
\end{eqnarray}
be imaginary. Since $g_{2}^{r}\Sigma_{0}^{<}g_{2}^{a}$ and $g^{r}\Sigma_{0}^{<}g^{a}$
are pure imaginary quantities, for $G_{\sigma\sigma}^{<}$ to be pure
imaginary, the real part of $A_{1}$ needs to cancel, i.e.,: 
\begin{equation}
Im\left(g^{r}\right)g_{2}^{r}\Sigma_{0}^{<}g_{2}^{a}=-Im\left(g_{2}^{a}\right)g^{r}\Sigma_{0}^{<}g^{a}.\label{eq:52}
\end{equation}
Define 
\begin{eqnarray}
\varepsilon_{0} & = & \varepsilon_{\sigma}+Re\left(\Sigma_{0}^{r}\right),\\
\Gamma & = & -Im\left(\Sigma_{0}^{r}\right),
\end{eqnarray}
and use it to rewrite the equation (\ref{eq:52}) as:
\begin{equation}
\frac{\Gamma}{\left(\hbar\omega-\varepsilon_{0}\right)^{2}+\left(\Gamma\right)^{2}}\Sigma_{0}^{<}\frac{1}{\left(\hbar\omega-\varepsilon_{0}-U\right)^{2}+\left(\Gamma\right)^{2}}=-\frac{-\Gamma}{\left(\hbar\omega-\varepsilon_{0}-U\right)^{2}+\left(\Gamma\right)^{2}}\Sigma_{0}^{<}\frac{1}{\left(\hbar\omega-\varepsilon_{0}\right)^{2}+\left(\Gamma\right)^{2}}.
\end{equation}
Obviously the real part of $A_{1}$ cancels, hence $G_{\sigma\sigma}^{<}$
is imaginary and fulfills the symmetry $G_{\sigma\sigma}^{<}\left(\omega\right)=-\left(G_{\sigma\sigma}^{<}\left(\omega\right)\right)^{*}$.

\section{Full derivation of the broken symmetry in the double Anderson model}

Now we refer to subsection \textbf{The double Anderson model} in the
manuscript. In what follows we show in greater detail that $\left(G_{\alpha\beta}^{\sigma\sigma}\left(\omega\right)\right)^{r}\neq\left(\left(G_{\beta\alpha}^{\sigma\sigma}\left(\omega\right)\right)^{a}\right)^{*}$.
Again, following the derivation in Ref.~\onlinecite{Joshi2000} we
define the following contour ordered GFs:
\begin{equation}
G_{\alpha\beta}^{\sigma\sigma}\left(t,t'\right)=-\frac{i}{\hbar}\left\langle T_{C}d_{\alpha\sigma}\left(t\right)d_{\beta\sigma}^{\dagger}\left(t'\right)\right\rangle ,
\end{equation}
\begin{equation}
\mathbb{\mathbb{G}_{\alpha\beta\gamma}^{\tau\sigma\sigma}}\left(t,t'\right)=-\frac{i}{\hbar}\left\langle T_{C}n_{\alpha\tau}\left(t\right)d_{\beta\sigma}\left(t\right)d_{\gamma\sigma}^{\dagger}\left(t'\right)\right\rangle .
\end{equation}
where $\tau=\sigma/\bar{\sigma}$ . The resulting EOM are:
\begin{eqnarray}
G_{\alpha\beta}^{\sigma\sigma}\left(\omega\right) & = & \left(\hbar\omega-\varepsilon_{\alpha,\sigma}-\Sigma_{0}\left(\omega\right)\right)^{-1}\times\left(\delta_{\alpha\beta}^{\sigma\sigma}+h_{\alpha\beta}^{\sigma}G_{\beta\beta}^{\sigma\sigma}\left(\omega\right)+U_{\alpha}\mathbb{G}_{\alpha\alpha\beta}^{\bar{\sigma}\sigma\sigma}\left(\omega\right)\right.\nonumber \\
 &  & \left.+V_{\alpha\beta}^{\sigma\bar{\sigma}}\mathbb{G}_{\beta\alpha\beta}^{\bar{\sigma}\sigma\sigma}\left(\omega\right)+V_{\alpha\beta}^{\sigma\sigma}\mathbb{G}_{\beta\alpha\beta}^{\sigma\sigma\sigma}\left(\omega\right)\right),\label{eq:Single part-1}
\end{eqnarray}
\begin{eqnarray}
\mathbb{G}_{\alpha\alpha\beta}^{\bar{\sigma}\sigma\sigma}\left(\omega\right) & = & \left(\hbar\omega-\varepsilon_{\alpha\sigma}-U_{\alpha}-V_{\alpha\beta}^{\sigma\sigma}\left\langle n_{\beta\sigma}\right\rangle -V_{\alpha\beta}^{\sigma\bar{\sigma}}\left\langle n_{\beta\bar{\sigma}}\right\rangle -\Sigma_{0}\left(\omega\right)\right)^{-1}\nonumber \\
 &  & \times\left[h_{\alpha\beta}^{\sigma}\mathbb{G}_{\alpha\beta\beta}^{\bar{\sigma}\sigma\sigma}\left(\omega\right)+\left\langle n_{\alpha\bar{\sigma}}\right\rangle \left(V_{\alpha\beta}^{\sigma\sigma}\mathbb{G}_{\beta\alpha\beta}^{\sigma\sigma\sigma}\left(\omega\right)+V_{\alpha\beta}^{\sigma\bar{\sigma}}\mathbb{G}_{\beta\alpha\beta}^{\bar{\sigma}\sigma\sigma}\left(\omega\right)\right)\right],\nonumber \\
\mathbb{G}_{\alpha\beta\beta}^{\bar{\sigma}\sigma\sigma}\left(\omega\right) & = & \left(\hbar\omega-\varepsilon_{\beta\sigma}-U_{\beta}\left\langle n_{\beta\bar{\sigma}}\right\rangle -V_{\beta\alpha}^{\sigma\sigma}\left\langle n_{\alpha\sigma}\right\rangle -V_{\beta\alpha}^{\sigma\bar{\sigma}}-\Sigma_{0}\left(\omega\right)\right)^{-1}\nonumber \\
 &  & \times\left[\left\langle n_{\alpha\bar{\sigma}}\right\rangle +h_{\beta\alpha}^{\sigma}\mathbb{G}_{\alpha\alpha\beta}^{\bar{\sigma}\sigma\sigma}\left(\omega\right)+\left\langle n_{\alpha\bar{\sigma}}\right\rangle \left(U_{\beta}\mathbb{G}_{\beta\beta\beta}^{\bar{\sigma}\sigma\sigma}\left(\omega\right)+V_{\beta\alpha}^{\sigma\sigma}\mathbb{G}_{\alpha\beta\beta}^{\sigma\sigma\sigma}\left(\omega\right)\right)\right],\nonumber \\
\mathbb{G}_{\alpha\beta\beta}^{\sigma\sigma\sigma}\left(\omega\right) & = & \left(\hbar\omega-\varepsilon_{\beta\sigma}-U_{\beta}\left\langle n_{\beta\bar{\sigma}}\right\rangle -V_{\beta\alpha}^{\sigma\bar{\sigma}}\left\langle n_{\alpha\bar{\sigma}}\right\rangle -V_{\beta\alpha}^{\sigma\sigma}-\Sigma_{0}\left(\omega\right)\right)^{-1}\nonumber \\
 &  & \times\left[\left\langle n_{\alpha\sigma}\right\rangle +h_{\beta\alpha}^{\sigma}\mathbb{G}_{\beta\alpha\beta}^{\sigma\sigma\sigma}\left(\omega\right)+\left\langle n_{\alpha\sigma}\right\rangle \left(U_{\beta}\mathbb{G}_{\beta\beta\beta}^{\bar{\sigma}\sigma\sigma}\left(\omega\right)+V_{\beta\alpha}^{\sigma\bar{\sigma}}\mathbb{G}_{\alpha\beta\beta}^{\bar{\sigma}\sigma\sigma}\left(\omega\right)\right)\right],\nonumber \\
\mathbb{G}_{\beta\alpha\beta}^{\bar{\sigma}\sigma\sigma}\left(\omega\right) & = & \left(\hbar\omega-\varepsilon_{\alpha\sigma}-U_{\alpha}\left\langle n_{\alpha\bar{\sigma}}\right\rangle -V_{\alpha\beta}^{\sigma\sigma}\left\langle n_{\beta\sigma}\right\rangle -V_{\alpha\beta}^{\sigma\sigma}-\Sigma_{0}\left(\omega\right)\right)^{-1}\\
 &  & \times\left[h_{\alpha\beta}^{\sigma}\mathbb{G}_{\beta\beta\beta}^{\bar{\sigma}\sigma\sigma}\left(\omega\right)+\left\langle n_{\beta\bar{\sigma}}\right\rangle \left(U_{\alpha}\mathbb{G}_{\alpha\alpha\beta}^{\bar{\sigma}\sigma\sigma}\left(\omega\right)+V_{\alpha\beta}^{\sigma,\sigma}\mathbb{G}_{\beta\alpha\beta}^{\sigma\sigma\sigma}\left(\omega\right)\right)\right],\nonumber \\
\mathbb{G}_{\beta\alpha\beta}^{\sigma\sigma\sigma}\left(\omega\right) & = & \left(\hbar\omega-\varepsilon_{\alpha\sigma}-U_{\alpha}\left\langle n_{\alpha\bar{\sigma}}\right\rangle -V_{\alpha\beta}^{\sigma\bar{\sigma}}\left\langle n_{\beta\bar{\sigma}}\right\rangle -V_{\alpha\beta}^{\sigma\sigma}-\Sigma_{0}\left(\omega\right)\right)^{-1}\nonumber \\
 &  & \times\left[-\left\langle d_{\beta\sigma}^{\dagger}d_{\alpha,\sigma}\right\rangle +h_{\alpha\beta}^{\sigma}\mathbb{G}_{\alpha\beta\beta}^{\sigma\sigma\sigma}\left(\omega\right)+\left\langle n_{\beta,\sigma}\right\rangle \left(U_{\alpha}\mathbb{G}_{\alpha\alpha\beta}^{\bar{\sigma}\sigma\sigma}\left(\omega\right)+V_{\alpha,\beta}^{\sigma,\bar{\sigma}}\mathbb{G}_{\beta\alpha\beta}^{\bar{\sigma}\sigma\sigma}\left(\omega\right)\right)\right],\nonumber \\
\mathbb{G}_{\beta\beta\beta}^{\bar{\sigma}\sigma\sigma}\left(\omega\right) & = & \left(\hbar\omega-\varepsilon_{\beta\sigma}-U_{\beta}-V_{\beta\alpha}^{\sigma\bar{\sigma}}\left\langle n_{\alpha\bar{\sigma}}\right\rangle -V_{\beta\alpha}^{\sigma\sigma}\left\langle n_{\alpha\sigma}\right\rangle -\Sigma_{0}\left(\omega\right)\right)^{-1}\nonumber \\
 &  & \times\left[\left\langle n_{\beta\bar{\sigma}}\right\rangle +h_{\beta\alpha}^{\sigma}\mathbb{G}_{\beta\alpha\beta}^{\bar{\sigma}\sigma\sigma}\left(\omega\right)+\left\langle n_{\beta\bar{\sigma}}\right\rangle \left(V_{\beta\alpha}^{\sigma,\sigma}\mathbb{G}_{\alpha\beta\beta}^{\sigma\sigma\sigma}\left(\omega\right)+V_{\beta\alpha}^{\sigma,\bar{\sigma}}\mathbb{G}_{\alpha\beta\beta}^{\bar{\sigma}\sigma\sigma}\left(\omega\right)\right)\right],\nonumber 
\end{eqnarray}
By applying the Langreth rules one can find the retarded and advanced
projections of the single particle GF (equation (\ref{eq:Single part-1})).
For simplicity we derive them for the case where $V_{ij}^{\sigma\tau}=0$.
Define (as usual omitting $\left(\omega\right)$ for brevity) 
\begin{eqnarray}
\left(g_{i}\right)^{r,a} & = & \frac{1}{\hbar\omega-\varepsilon_{i,\sigma}-\Sigma_{0}^{r,a}},\\
\left(g_{ii}^{\bar{\sigma}\sigma}\right)^{r,a} & = & \frac{1}{\hbar\omega-\varepsilon_{i,\sigma}-U_{i}-\Sigma_{0}^{r,a}},\\
\left(g_{ij}^{\bar{\sigma}\sigma}\right)^{r,a} & = & \frac{1}{\hbar\omega-\varepsilon_{j,\sigma}-U_{j}\left\langle n_{j,\bar{\sigma}}\right\rangle -\Sigma_{0}^{r,a}},
\end{eqnarray}
where $\Sigma_{0}$ is defined in equation (\ref{eq:selfenergy}).
Now we are ready to look at the equation we get for $\left(G_{\alpha\beta}^{\sigma\sigma}\left(\omega\right)\right)^{r}$. 

\begin{eqnarray}
\left(G_{\alpha\beta}^{\sigma\sigma}\right)^{r} & = & \left(g_{\alpha}\right)^{r}h_{\alpha,\beta}^{\sigma}\left(G_{\beta\beta}^{\sigma\sigma}\right)^{r}+\left(g_{\alpha}\right)^{r}U_{\alpha}\left(\mathbb{G}_{\alpha\alpha\beta}^{\bar{\sigma}\sigma\sigma}\right)^{r},
\end{eqnarray}
\begin{eqnarray}
\left(G_{\beta\beta}^{\sigma\sigma}\right)^{r} & = & \left(g_{\beta}\right)^{r}+\left(g_{\beta}\right)^{r}h_{\beta,\alpha}^{\sigma}\left(G_{\alpha\beta}^{\sigma\sigma}\right)^{r}+\left(g_{\beta}\right)^{r}U_{\beta}\left(\mathbb{G}_{\beta\beta\beta}^{\bar{\sigma}\sigma\sigma}\right)^{r},
\end{eqnarray}
\begin{eqnarray}
\left(\mathbb{G}_{\alpha\alpha\beta}^{\bar{\sigma}\sigma\sigma}\right)^{r} & = & \left(g_{\alpha\alpha}^{\bar{\sigma}\sigma}\right)^{r}h_{\alpha,\beta}^{\sigma}\left(\mathbb{G}_{\alpha\beta\beta}^{\bar{\sigma}\sigma\sigma}\right)^{r},\label{eq:1st}
\end{eqnarray}
\begin{eqnarray}
\left(\mathbb{G}_{\beta\alpha\beta}^{\bar{\sigma}\sigma\sigma}\right)^{r} & = & \left(g_{\beta\alpha}^{\bar{\sigma}\sigma}\right)^{r}h_{\alpha\beta}^{\sigma}\left(\mathbb{G}_{\beta\beta\beta}^{\bar{\sigma}\sigma\sigma}\right)^{r}+\left(g_{\beta\alpha}^{\bar{\sigma}\sigma}\right)^{r}\left\langle n_{\beta\bar{\sigma}}\right\rangle U_{\alpha}\left(\mathbb{G}_{\alpha\alpha\beta}^{\bar{\sigma}\sigma\sigma}\right)^{r},\label{eq:2nd}
\end{eqnarray}
\begin{eqnarray}
\left(\mathbb{G}_{\beta\beta\beta}^{\bar{\sigma}\sigma\sigma}\right)^{r} & = & \left(g_{\beta\beta}^{\bar{\sigma}\sigma}\right)^{r}\left\langle n_{\beta\bar{\sigma}}\right\rangle +\left(g_{\beta\beta}^{\bar{\sigma}\sigma}\right)^{r}h_{\beta\alpha}^{\sigma}\left(\mathbb{G}_{\beta\alpha\beta}^{\bar{\sigma}\sigma\sigma}\right)^{r},\label{eq:3rd}
\end{eqnarray}

\begin{eqnarray}
\left(\mathbb{G}_{\alpha\beta\beta}^{\bar{\sigma}\sigma\sigma}\right)^{r} & =\left(g_{\alpha\beta}^{\bar{\sigma}\sigma}\right)^{r}\left\langle n_{\alpha,\bar{\sigma}}\right\rangle + & \left(g_{\alpha\beta}^{\bar{\sigma}\sigma}\right)^{r}h_{\beta\alpha}^{\sigma}\left(\mathbb{G}_{\alpha\alpha\beta}^{\bar{\sigma}\sigma\sigma}\right)^{r}+\left(g_{\alpha\beta}^{\bar{\sigma}\sigma}\right)^{r}\left\langle n_{\alpha,\bar{\sigma}}\right\rangle U_{\beta}\left(\mathbb{G}_{\beta\beta\beta}^{\bar{\sigma}\sigma\sigma}\right)^{r}.\label{eq:4th}
\end{eqnarray}
Substituting $\left(G_{\beta\beta}^{\sigma\sigma}\right)^{r}$ into
the equation of $\left(G_{\alpha\beta}^{\sigma\sigma}\right)^{r}:$

\begin{eqnarray}
\left(G_{\alpha\beta}^{\sigma\sigma}\right)^{r} & = & \left(g_{\alpha}\right)^{r}h_{\alpha,\beta}^{\sigma}\left(\left(g_{\beta}\right)^{r}+\left(g_{\beta}\right)^{r}h_{\beta,\alpha}^{\sigma}\left(G_{\alpha\beta}^{\sigma\sigma}\right)^{r}+\left(g_{\beta}\right)^{r}U_{\beta}\left(\mathbb{G}_{\beta\beta\beta}^{\bar{\sigma}\sigma\sigma}\right)^{r}\right)+\left(g_{\alpha}\right)^{r}U_{\alpha}\left(\mathbb{G}_{\alpha\alpha\beta}^{\bar{\sigma}\sigma\sigma}\right)^{r},\nonumber \\
\end{eqnarray}
\begin{eqnarray}
\left(G_{\alpha\beta}^{\sigma\sigma}\right)^{r} & = & \left(I-\left(g_{\alpha}\right)^{r}h_{\alpha,\beta}^{\sigma}\left(g_{\beta}\right)^{r}h_{\beta,\alpha}^{\sigma}\right)^{-1}\nonumber \\
 &  & \times\left(\left(g_{\alpha}\right)^{r}h_{\alpha,\beta}^{\sigma}\left(g_{\beta}\right)^{r}+\left(g_{\alpha}\right)^{r}h_{\alpha,\beta}^{\sigma}\left(g_{\beta}\right)^{r}U_{\beta}\left(\mathbb{G}_{\beta\beta\beta}^{\bar{\sigma}\sigma\sigma}\right)^{r}+\left(g_{\alpha}\right)^{r}U_{\alpha}\left(\mathbb{G}_{\alpha\alpha\beta}^{\bar{\sigma}\sigma\sigma}\right)^{r}\right).\label{eq:Grfinal}
\end{eqnarray}
Now we need the equations for $\left(\mathbb{G}_{\alpha\alpha\beta}^{\bar{\sigma}\sigma\sigma}\right)^{r}$
and $\left(\mathbb{G}_{\beta\beta\beta}^{\bar{\sigma}\sigma\sigma}\right)^{r}$.
Using equations (\ref{eq:1st}) to (\ref{eq:4th}) we get:

\begin{eqnarray}
\left(\mathbb{G}_{\alpha\alpha\beta}^{\bar{\sigma}\sigma\sigma}\right)^{r} & = & \left(1-\left(g_{\alpha\alpha}^{\bar{\sigma}\sigma}\right)^{r}h_{\alpha,\beta}^{\sigma}\left(g_{\alpha\beta}^{\bar{\sigma}\sigma}\right)^{r}h_{\beta\alpha}^{\sigma}-\left(g_{\alpha\alpha}^{\bar{\sigma}\sigma}\right)^{r}h_{\alpha,\beta}^{\sigma}\left(g_{\alpha\beta}^{\bar{\sigma}\sigma}\right)^{r}\left\langle n_{\alpha,\bar{\sigma}}\right\rangle U_{\beta}\right.\nonumber \\
 &  & \times\left.\left(1-\left(g_{\beta\beta}^{\bar{\sigma}\sigma}\right)^{r}h_{\beta\alpha}^{\sigma}\left(g_{\beta\alpha}^{\bar{\sigma}\sigma}\right)^{r}h_{\alpha\beta}^{\sigma}\right)^{-1}\left(g_{\beta\beta}^{\bar{\sigma}\sigma}\right)^{r}h_{\beta\alpha}^{\sigma}\left(g_{\beta\alpha}^{\bar{\sigma}\sigma}\right)^{r}\left\langle n_{\beta\bar{\sigma}}\right\rangle U_{\alpha}\right)^{-1}\nonumber \\
 &  & \times\left(\left(g_{\alpha\alpha}^{\bar{\sigma}\sigma}\right)^{r}h_{\alpha,\beta}^{\sigma}\left(g_{\alpha\beta}^{\bar{\sigma}\sigma}\right)^{r}\left\langle n_{\alpha,\bar{\sigma}}\right\rangle +\left(g_{\alpha\alpha}^{\bar{\sigma}\sigma}\right)^{r}h_{\alpha,\beta}^{\sigma}\left(g_{\alpha\beta}^{\bar{\sigma}\sigma}\right)^{r}\left\langle n_{\alpha,\bar{\sigma}}\right\rangle U_{\beta}\right.\nonumber \\
 &  & \times\left.\left(1-\left(g_{\beta\beta}^{\bar{\sigma}\sigma}\right)^{r}h_{\beta\alpha}^{\sigma}\left(g_{\beta\alpha}^{\bar{\sigma}\sigma}\right)^{r}h_{\alpha\beta}^{\sigma}\right)^{-1}\left(g_{\beta\beta}^{\bar{\sigma}\sigma}\right)^{r}\left\langle n_{\beta\bar{\sigma}}\right\rangle \right),
\end{eqnarray}
and 

\begin{eqnarray}
\left(\mathbb{G}_{\beta\beta\beta}^{\bar{\sigma}\sigma\sigma}\right)^{r} & = & \left(1-\left(g_{\beta\beta}^{\bar{\sigma}\sigma}\right)^{r}h_{\beta\alpha}^{\sigma}\left(g_{\beta\alpha}^{\bar{\sigma}\sigma}\right)^{r}h_{\alpha\beta}^{\sigma}-\left(g_{\beta\beta}^{\bar{\sigma}\sigma}\right)^{r}h_{\beta\alpha}^{\sigma}\left(g_{\beta\alpha}^{\bar{\sigma}\sigma}\right)^{r}\left\langle n_{\beta\bar{\sigma}}\right\rangle U_{\alpha}\right.\nonumber \\
 &  & \times\left.\left(1-\left(g_{\alpha\alpha}^{\bar{\sigma}\sigma}\right)^{r}h_{\alpha,\beta}^{\sigma}\left(g_{\alpha\beta}^{\bar{\sigma}\sigma}\right)^{r}h_{\beta\alpha}^{\sigma}\right)^{-1}\left(g_{\alpha\alpha}^{\bar{\sigma}\sigma}\right)^{r}h_{\alpha,\beta}^{\sigma}\left(g_{\alpha\beta}^{\bar{\sigma}\sigma}\right)^{r}\left\langle n_{\alpha,\bar{\sigma}}\right\rangle U_{\beta}\right)^{-1}\nonumber \\
 &  & \times\left(\left(g_{\beta\beta}^{\bar{\sigma}\sigma}\right)^{r}\left\langle n_{\beta\bar{\sigma}}\right\rangle +\left(g_{\beta\beta}^{\bar{\sigma}\sigma}\right)^{r}h_{\beta\alpha}^{\sigma}\left(g_{\beta\alpha}^{\bar{\sigma}\sigma}\right)^{r}\left\langle n_{\beta\bar{\sigma}}\right\rangle U_{\alpha}\right.\nonumber \\
 &  & \times\left.\left(1-\left(g_{\alpha\alpha}^{\bar{\sigma}\sigma}\right)^{r}h_{\alpha,\beta}^{\sigma}\left(g_{\alpha\beta}^{\bar{\sigma}\sigma}\right)^{r}h_{\beta\alpha}^{\sigma}\right)^{-1}\left(g_{\alpha\alpha}^{\bar{\sigma}\sigma}\right)^{r}h_{\alpha,\beta}^{\sigma}\left(g_{\alpha\beta}^{\bar{\sigma}\sigma}\right)^{r}\left\langle n_{\alpha,\bar{\sigma}}\right\rangle \right).
\end{eqnarray}
The same way one can derive an expression for $\left(G_{\beta\alpha}^{\sigma\sigma}\right)^{a}$
\begin{eqnarray}
\left(G_{\beta\alpha}^{\sigma\sigma}\right)^{a} & = & \left(I-\left(g_{\beta}\right)^{a}h_{\beta,\alpha}^{\sigma}\left(g_{\alpha}\right)^{a}h_{\alpha,\beta}^{\sigma}\right)^{-1}\nonumber \\
 &  & \times\left(\left(g_{\beta}\right)^{a}h_{\beta,\alpha}^{\sigma}\left(g_{\alpha}\right)^{a}+\left(g_{\beta}\right)^{a}h_{\beta,\alpha}^{\sigma}\left(g_{\alpha}\right)^{a}U_{\alpha}\left(\mathbb{G}_{\alpha\alpha\alpha}^{\bar{\sigma}\sigma\sigma}\right)^{a}+\left(g_{\beta}\right)^{a}U_{\beta}\left(\mathbb{G}_{\beta\beta\alpha}^{\bar{\sigma}\sigma\sigma}\right)^{a}\right),\label{eq:Gafinal}
\end{eqnarray}
with
\begin{eqnarray}
\left(\mathbb{G}_{\beta\beta\alpha}^{\bar{\sigma}\sigma\sigma}\right)^{a} & = & \left(1-\left(g_{\beta\beta}^{\bar{\sigma}\sigma}\right)^{a}h_{\beta,\alpha}^{\sigma}\left(g_{\beta\alpha}^{\bar{\sigma}\sigma}\right)^{a}h_{\alpha\beta}^{\sigma}-\left(g_{\beta\beta}^{\bar{\sigma}\sigma}\right)^{a}h_{\beta,\alpha}^{\sigma}\left(g_{\beta\alpha}^{\bar{\sigma}\sigma}\right)^{a}\left\langle n_{\beta,\bar{\sigma}}\right\rangle U_{\alpha}\right.\nonumber \\
 &  & \times\left.\left(1-\left(g_{\alpha\alpha}^{\bar{\sigma}\sigma}\right)^{a}h_{\alpha\beta}^{\sigma}\left(g_{\alpha\beta}^{\bar{\sigma}\sigma}\right)^{a}h_{\beta\alpha}^{\sigma}\right)^{-1}\left(g_{\alpha\alpha}^{\bar{\sigma}\sigma}\right)^{a}h_{\alpha\beta}^{\sigma}\left(g_{\alpha\beta}^{\bar{\sigma}\sigma}\right)^{a}\left\langle n_{\alpha\bar{\sigma}}\right\rangle U_{\beta}\right)^{-1}\nonumber \\
 &  & \times\left(\left(g_{\beta\beta}^{\bar{\sigma}\sigma}\right)^{a}h_{\beta,\alpha}^{\sigma}\left(g_{\beta\alpha}^{\bar{\sigma}\sigma}\right)^{a}\left\langle n_{\beta,\bar{\sigma}}\right\rangle +\left(g_{\beta\beta}^{\bar{\sigma}\sigma}\right)^{a}h_{\beta,\alpha}^{\sigma}\left(g_{\beta\alpha}^{\bar{\sigma}\sigma}\right)^{a}\left\langle n_{\beta,\bar{\sigma}}\right\rangle U_{\alpha}\right.\nonumber \\
 &  & \times\left.\left(1-\left(g_{\alpha\alpha}^{\bar{\sigma}\sigma}\right)^{a}h_{\alpha\beta}^{\sigma}\left(g_{\alpha\beta}^{\bar{\sigma}\sigma}\right)^{a}h_{\beta\alpha}^{\sigma}\right)^{-1}\left(g_{\alpha\alpha}^{\bar{\sigma}\sigma}\right)^{a}\left\langle n_{\alpha\bar{\sigma}}\right\rangle \right),
\end{eqnarray}
and 
\begin{eqnarray}
\left(\mathbb{G}_{\alpha\alpha\alpha}^{\bar{\sigma}\sigma\sigma}\right)^{a} & = & \left(1-\left(g_{\alpha\alpha}^{\bar{\sigma}\sigma}\right)^{a}h_{\alpha\beta}^{\sigma}\left(g_{\alpha\beta}^{\bar{\sigma}\sigma}\right)^{a}h_{\beta\alpha}^{\sigma}-\left(g_{\alpha\alpha}^{\bar{\sigma}\sigma}\right)^{a}h_{\alpha\beta}^{\sigma}\left(g_{\alpha\beta}^{\bar{\sigma}\sigma}\right)^{a}\left\langle n_{\alpha\bar{\sigma}}\right\rangle U_{\beta}\right.\nonumber \\
 &  & \times\left.\left(1-\left(g_{\beta\beta}^{\bar{\sigma}\sigma}\right)^{a}h_{\beta,\alpha}^{\sigma}\left(g_{\beta\alpha}^{\bar{\sigma}\sigma}\right)^{a}h_{\alpha\beta}^{\sigma}\right)^{-1}\left(g_{\beta\beta}^{\bar{\sigma}\sigma}\right)^{a}h_{\beta,\alpha}^{\sigma}\left(g_{\beta\alpha}^{\bar{\sigma}\sigma}\right)^{a}\left\langle n_{\beta,\bar{\sigma}}\right\rangle U_{\alpha}\right)^{-1}\nonumber \\
 &  & \times\left(\left(g_{\alpha\alpha}^{\bar{\sigma}\sigma}\right)^{a}\left\langle n_{\alpha\bar{\sigma}}\right\rangle +\left(g_{\alpha\alpha}^{\bar{\sigma}\sigma}\right)^{a}h_{\alpha\beta}^{\sigma}\left(g_{\alpha\beta}^{\bar{\sigma}\sigma}\right)^{a}\left\langle n_{\alpha\bar{\sigma}}\right\rangle U_{\beta}\right.\nonumber \\
 &  & \times\left.\left(1-\left(g_{\beta\beta}^{\bar{\sigma}\sigma}\right)^{a}h_{\beta,\alpha}^{\sigma}\left(g_{\beta\alpha}^{\bar{\sigma}\sigma}\right)^{a}h_{\alpha\beta}^{\sigma}\right)^{-1}\left(g_{\beta\beta}^{\bar{\sigma}\sigma}\right)^{a}h_{\beta,\alpha}^{\sigma}\left(g_{\beta\alpha}^{\bar{\sigma}\sigma}\right)^{a}\left\langle n_{\beta,\bar{\sigma}}\right\rangle \right).
\end{eqnarray}
The question we now ask is whether 
\begin{equation}
\left(G_{\alpha\beta}^{\sigma\sigma}\right)^{r}=\left(\left(G_{\beta\alpha}^{\sigma\sigma}\right)^{a}\right)^{*}.
\end{equation}
Using
\begin{equation}
\left(g_{i}\right)^{r}=\left(\left(g_{i}\right)^{a}\right)^{*},\quad\left(g_{\bar{i}i}^{2}\right)^{r}=\left(\left(g_{\bar{i}i}^{2}\right)^{a}\right)^{*},\quad\left(g_{\bar{i}j}^{2}\right)^{r}=\left(\left(g_{\bar{i}j}^{2}\right)^{a}\right)^{*},
\end{equation}
and looking back at equations (\ref{eq:Grfinal}) and (\ref{eq:Gafinal})
we find that 
\begin{equation}
\left(I-\left(g_{\alpha}\right)^{r}h_{\alpha,\beta}^{\sigma}\left(g_{\beta}\right)^{r}h_{\beta,\alpha}^{\sigma}\right)^{-1}=\left(\left(I-\left(g_{\beta}\right)^{a}h_{\beta,\alpha}^{\sigma}\left(g_{\alpha}\right)^{a}h_{\alpha,\beta}^{\sigma}\right)^{-1}\right)^{*},
\end{equation}
and that 
\begin{equation}
\left(g_{\alpha}\right)^{r}h_{\alpha,\beta}^{\sigma}\left(g_{\beta}\right)^{r}=\left(\left(g_{\beta}\right)^{a}h_{\beta,\alpha}^{\sigma}\left(g_{\alpha}\right)^{a}\right)^{*}.
\end{equation}
Therefore, it is sufficient check whether the next equality 
\begin{eqnarray}
\left(g_{\alpha}\right)^{r}U_{\alpha}\left(\mathbb{G}_{\alpha\alpha\beta}^{\bar{\sigma}\sigma\sigma}\right)^{r}+\left(g_{\alpha}\right)^{r}h_{\alpha,\beta}^{\sigma}\left(g_{\beta}\right)^{r}U_{\beta}\left(\mathbb{G}_{\beta\beta\beta}^{\bar{\sigma}\sigma\sigma}\right)^{r} & = & \left(\left(g_{\beta}\right)^{a}h_{\beta,\alpha}^{\sigma}\left(g_{\alpha}\right)^{a}U_{\alpha}\left(\mathbb{G}_{\alpha\alpha\alpha}^{\bar{\sigma}\sigma\sigma}\right)^{a}+\left(g_{\beta}\right)^{a}U_{\beta}\left(\mathbb{G}_{\beta\beta\alpha}^{\bar{\sigma}\sigma\sigma}\right)^{a}\right)^{*},\nonumber \\
\label{eq:last-1}
\end{eqnarray}
holds. Substitute the equations for $\left(\mathbb{G}_{\alpha\alpha\beta}^{\bar{\sigma}\sigma\sigma}\right)^{r}$,
$\left(\mathbb{G}_{\beta\beta\beta}^{\bar{\sigma}\sigma\sigma}\right)^{r}$,
$\left(\mathbb{G}_{\alpha\alpha\alpha}^{\bar{\sigma}\sigma\sigma}\right)^{a}$
and $\left(\mathbb{G}_{\beta\beta\alpha}^{\bar{\sigma}\sigma\sigma}\right)^{a}$
into equation (\ref{eq:last-1}) and after some tedious algebra we
find that unless 
\begin{equation}
\left(g_{\alpha\beta}^{\bar{\sigma}\sigma}\right)^{r}=\left(\left(g_{\beta}\right)^{a}\right)^{*}=\left(g_{\beta}\right)^{r},
\end{equation}
the identity $\left(G_{\alpha\beta}^{\sigma\sigma}\right)^{r}=\left(\left(G_{\beta\alpha}^{\sigma\sigma}\right)^{a}\right)^{*}$does
not hold.\textbf{ }But as 
\begin{equation}
\left(g_{\alpha\beta}^{\bar{\sigma}\sigma}\right)^{r}=\frac{1}{\hbar\omega-\varepsilon_{\beta,\sigma}-U_{\beta}\left\langle n_{\beta,\bar{\sigma}}\right\rangle -\Sigma_{0}^{r}},
\end{equation}
and 
\begin{equation}
\left(g_{\beta}\right)^{r}=\frac{1}{\hbar\omega-\varepsilon_{\beta,\sigma}-\Sigma_{0}^{r}},
\end{equation}
it is obvious that $\left(g_{\alpha\beta}^{\bar{\sigma}\sigma}\right)^{r}\neq\left(\left(g_{\beta}\right)^{a}\right)^{*}$,
hence finally $\left(G_{\alpha\beta}^{\sigma\sigma}\right)^{r}\neq\left(\left(G_{\beta\alpha}^{\sigma\sigma}\right)^{a}\right)^{*}$.
The same can be done to show that $\left(G_{\alpha\beta}^{\sigma\sigma}\right)^{<,>}\neq-\left(\left(G_{\beta\alpha}^{\sigma\sigma}\right)^{<,>}\right)^{*}$
and $\left(G_{\alpha\beta}^{\sigma\sigma}\right)^{r}-\left(G_{\alpha\beta}^{\sigma\sigma}\right)^{a}\neq\left(G_{\alpha\beta}^{\sigma\sigma}\right)^{>}-\left(G_{\alpha\beta}^{\sigma\sigma}\right)^{<}.$

\end{widetext}
\end{document}